\documentclass[11pt,eqsecnum]{article}

\usepackage{amsmath, amssymb}
 \usepackage{epsfig}
\usepackage{mathrsfs}
\usepackage{psfrag}
\usepackage{graphicx}
\usepackage{epstopdf}

\date{}

\numberwithin{equation}{section}

\newcommand{\E}{{\cal E}}
\newcommand{\J}{{\cal J}}

\newcommand{\EE}{{\mathbb{E}}}
\newcommand{\KK}{{\mathbb{K}}}

\def\l{\lambda}
\def\sl{\sqrt{\lambda}}
\def \ov {\over}
\def\k{\kappa}
\def\g{\gamma}
\def \tS{\bar {\cal S}}
\def\bi{\bibitem}
\def\hS{{\hat \cS}}
 \def\f{\rm f}
 \def\tf{\tilde{\rm f}}
\def\tg{\tilde\gamma}
\def\ci{\cite}
\def\foot{\footnote}
\def\ha{\frac{1}{2}}
\def\PP{\f}
\newcommand{\rf}[1]{(\ref{#1})}
\def\bea{\begin{eqnarray}}
\def\eea{\end{eqnarray}}

\def\no{\nonumber}

\def\C{{\cal C}}

\newcommand{\superN}{\mathcal{N}}

\newcommand{\gym}{g\inddowns{YM}}

\newcommand{\inddowns}[1]{_{\mathrm{\scriptscriptstyle #1}}}

\newcommand{\be}{\begin{equation}}
\newcommand{\ee}{\end{equation}}
\newcommand{\ba}{\begin{eqnarray}}
\newcommand{\ea}{\end{eqnarray}}

\newcommand{\refeq}[1]{Eq.~(\ref{eq:#1})}

\newcommand{\ads}{AdS_5\times S^5}

\def\cS{{\cal S}}

\def\cP{{\cal P}}

\def\vare{\varepsilon}

\newcommand{\Mellin}[1]{{\cal M}\left[#1\right]}

\allowdisplaybreaks

\usepackage{color}

\begin{document}

\title{\hfill {\small AEI-2010-017}\vskip 1cm
Generalized Gribov-Lipatov Reciprocity and AdS/CFT}

\author{Matteo Beccaria$^{a}$\footnote{beccaria@le.infn.it}, \ Valentina Forini$^{b}$\footnote{forini@aei.mpg.de} \  and Guido Macorini$^{a}$\footnote{guido.macorini@le.infn.it}}

\maketitle

\begin{center}
\emph{
$^{a}$
Physics Department, Salento University and INFN, 73100 Lecce, Italy \\
\vskip 0.04cm
$^{b}$ Max-Planck-Institut f\"ur Gravitationsphysik,  Albert-Einstein-Institut \\
Am M\"uhlenberg 1, D-14476 
Potsdam, Germany
 }
\end{center}

\begin{abstract}
 Planar $\mathcal{N}=4$ SYM theory and QCD share the gluon sector, suggesting the investigation of 
Gribov-Lipatov reciprocity in the supersymmetric theory. Since  the
AdS/CFT correspondence links $\mathcal{N}=4$ SYM and superstring dynamics on $\ads$, 
reciprocity is also expected to show up in the quantum corrected energies of certain classical string configurations
dual to gauge theory twist-operators. We review recent results confirming this picture and 
revisiting the old idea of Gribov-Lipatov reciprocity as a modern theoretical tool useful for the study of open problems in AdS/CFT.
\end{abstract}

\tableofcontents

\section{Introduction and overview}

An intense activity in the study of the duality between the planar, large N limit of the $\mathcal{N}=4$ super-Yang-Mills (SYM) theory with $SU(N)$ gauge group and the free type IIB superstrings in $AdS_5\times S^5$ is based on the development of analytic tools that exploit the classical integrability of the string side~\cite{Bena:2003wd},  as well as an {\em   internal} integrability of the superconformal theory~\cite{Beisert:2004ry}. In the latter case, the scale dependence of renormalized composite operators is governed, even at higher loops, by a local, integrable, super spin chain Hamiltonian whose interaction range increases with the loop order~\cite{Beisert:2003yb,Beisert:2005fw}. This fact has firstly set  the long range asymptotic Bethe equations of ~\cite{Beisert:2005fw} as a natural tool for calculating anomalous dimensions of the gauge single traces operators of the theory.  Although the relevant two-particle scattering matrix~\cite{Staudacher:2004tk}
   was determined in a gauge theory framework~\cite{Beisert:2005tm}, its tensor structure agrees with perturbative calculations in the gauge-fixed world-sheet theory~\cite{Klose:2006zd}.
  Its form is determined by the global symmetry of the two  theories, $\mathfrak{psu}(2,2|4)$, up to a phase (dressing factor) for which a crossing-like equation has been proposed~\cite{Janik:2006dc}.
 For its solution~\cite{Hernandez:2006tk}, based  also on 1-loop string data~\cite{BeisTsey}, an analytically continued weak-coupling expansion has been formulated~\cite{Beisert:2006ez}, whose effects on the anomalous dimensions of the twist-two operators remarkably agree with the direct calculation of the four-loop cusp anomalous dimension~\cite{Bern:2006ew}.
As a result, from the asymptotic Bethe equations (ABA) an integral equation for such cusp anomaly (or universal scaling function) has been derived, on which in fact is based  one of the most non-trivial tests of the structure of the AdS/CFT correspondence. Its strong coupling solution~\cite{bassocusp} (see also~\cite{StrongSL2}) has been in fact shown to perfectly match the expression for the cusp anomaly up to 2-loops term as computed directly from the quantum superstring~\cite{RoibanTseCusp}.

Due to their asymptotic nature, the Bethe equations furnish predictions for the anomalous dimensions that, for ''short'' operators~\cite{Kotikov:2007cy}, need to be corrected  by wrapping effects \cite{Ambjorn:2005wa}.  To this aim, a clever generalization of the L\"uscher 
formulas \cite{Luscher:1985dn} has
successfully given the correct finite-size correction in~\cite{Janikprev,Bajnok:2008bm} 
to the asymptotic anomalous dimension derived from the Bethe Ansatz~\cite{Kotikov:2007cy}, which has been confirmed by a purely field-theoretical calculation \cite{Fiamberti:2007rj}.
For the complete spectral equations of $\superN=4$ SYM, however, it is believed that thermodynamic Bethe Ansatz (TBA) methods ought to be applied, as has been initiated conjecturing a Y-system, which should yield anomalous dimensions of arbitrary local operators of planar $\superN=4$ SYM~\cite{Ysystem}, and TBA equations for string and gauge theory~\cite{TBA}. 
Relevant tests of these proposals have been already carried on~\cite{BFLZ,Gromov:2009zb,Gromov:2009tq}, which however, in the case of short operators anomalous dimensions at strong coupling~\cite{Gromov:2009zb},  still have to find a full numerical agreement with purely string theoretical computations~\cite{Arutyunov:2005hd,roibankonishi}~\footnote{The purely field-theoretical predictions in~\cite{Arutyunov:2005hd} and~\cite{roibankonishi} on the strong coupling expansion of the anomalous dimension for the Konishi operator differ both from~\cite{Gromov:2009zb} as well as  from each other.} and might need further elaboration~\cite{Rej:2009dk,Arutyunov:2009ax}.

To the purpose of furnishing  closed formulas for anomalous dimensions
which might check the TBA proposals at high orders of perturbation theory, the asymptotic Bethe equations, corrected with generalized L\"uscher formulas and further inputs, still stand  as a powerful tool for multi-loop calculations~\cite{BFLZ,5loopstwist2}.
The class of operators mostly relevant in this framework are the twist operators, also named  \emph{quasipartonic} ~\cite{quasipart}. These are single trace operators constructed with an arbitrary number of light-cone derivatives acting on the fundamental fields (scalars, gauginos or gauge fields). Their anomalous dimension depends on their spin (total number of derivatives), and their interest relies on their similarities with the QCD  twist operators arising in the analysis of deep inelastic scattering~\cite{Altarelli:1981ax}. 

It is a general fact that, while ${\cal N}=4$ SYM and QCD are in many details different, a compared analysis of  their properties has been crucial for a deeper understanding of  both of them. Integrability itself appeared for the first time in four-dimensional gauge field theories
in a QCD context,  in the high-energy Regge behavior of scattering amplitudes and in the scale dependence of composite  operators~\cite{QCD-int}. About conformal symmetry, unbroken in QCD at one loop, it does  not appear to be a necessary condition for  integrability, as discussed in~\cite{Belitsky:2004sc,Belitsky:2004sf,Belitsky:2005bu,DiVecchia:2004jw}, but it certainly plays an important role by imposing selection rules and multiplet structures. A notable common issue between ${\cal N}=4$ SYM  and QCD is their infrared structure~\cite{dixon}, and it is believed that QCD would benefit a lot from an ultimate all-loop solution of its superconformal version, since this would provide a representation for the ``dominant'' part of the perturbative gluon dynamics~\cite{Dokshitzer:2008zz}.

A remarkable example of such an interplay between $\mathcal{N}=4$ SYM theory and QCD in the framework of integrability is the  \emph{maximum transcendentality principle}~\cite{MTP}, according to which the anomalous dimension of twist-two operators at $n$ loops is a linear combination of generalized  harmonic sums of transcendentality $2n-1$. 
The principle has been the key via which closed multi-loop expressions for the anomalous dimension of special twist operators have been derived~\cite{Staudacher:2004tk,Kotikov:2007cy,Beccaria:2007cn,Beccaria:2007vh,Beccaria:2007bb,Beccaria:2007pb,Beccaria:2007gu,
BFLZ,5loopstwist2} and has been independently confirmed in a space-time framework~\cite{Bern:2005iz} as well as  exploiting the  Baxter approach~\cite{Kotikov:2008pv}. 
A second crucial connection is the relationship to the Balitsky-Fadin-Kuraev-Lipatov (BFKL) approach~\cite{BFKL} for describing high energy scattering amplitudes in gauge theory, 
which furnishes a  prediction  for the pole structure of the analytically continued anomalous dimensions of   twist operators.
The (supersymmetric generalization of the) BFKL equation appears to be a testing device for any conjecture on the exact higher-loop spectrum  of anomalous dimensions in the $\mathcal{N} = 4$ model, and in fact it  was determinant to state both the failure of Bethe equations in describing the spectrum of short operators~\cite{Kotikov:2007cy} as well as the correctness of the full result including the wrapping correction~\cite{Bajnok:2008qj}.

\bigskip

In this Review we will report on another fascinating and as yet not fully explained link
between QCD, $\mathcal{N}=4$ SYM and string theory. This is centered on the so-called \emph{reciprocity}, and consists in a surprising pattern that emerges in studying all the available anomalous dimensions of twist-two operators in QCD, their analogue in N = 4 SYM together with the energies of their dual string configurations.
The reciprocity condition is a constraint on the large spin behavior of a transform of the anomalous dimension, which should run in even negative powers of the Casimir of the collinear group $SL(2;\mathbb{R})$. This constraint, arising in the QCD context, has been presented in~\cite{dok1,dok2} as a special (space-time symmetric) reformulation of the parton distribution function evolution equations, while  in~\cite{bk} it has been approached from the point of view of the large spin expansion and  generalized to operators of arbitrary twist.  Reciprocity  has been checked in various multi-loop calculations of weakly coupled  ${\cal N}=4$ gauge theory~\cite{Beccaria:2008fi,Forini:2008ky,bdm,Beccaria:2007pb,Beccaria:2007cn,Beccaria:2009vt}. 

The AdS/CFT correspondence is the natural tool to investigate the presence of reciprocity relations at strong coupling. Since the planar perturbation theory should be convergent, such an organized structure of subleading terms in the large spin expansion should be visible also in the energies of the semiclassical string states corresponding to twist operators. Such strong coupling analysis, initiated in~\cite{bk} for a particular solution at the classical level, has been extended in ~\cite{BFTT} to more general configurations and beyond the classical result. 
 Given the complicated form of the relevant solutions, however, the large spin expansion for  corrections to the leading string energy is a non trivial task.  Remarkably, although not as manifestly as in the weak coupling case, also here the underlying integrable structure of the AdS/CFT system plays a crucial role in making feasible the analysis of reciprocity. The recent findings of~\cite{BDFPT}, demonstrating that the semi-classical fluctuation problem is governed by simple finite-gap operators, has provided us with analytic expressions for the fluctuation determinants that permit to carry out well-defined expansions in the large spin limit.
As a notable outcome, the large spin expansion of the string energy happens to have exactly the same structure as that of the anomalous dimension in the perturbative gauge theory, respecting reciprocity relations up to one-loop in string perturbation theory. 
Interesting generalizations of this analysis at strong coupling are the study~\cite{Beccaria:2009yt} of reciprocity for the for the first commuting charges defined in~\cite{Arutyunov:2003rg}, as well as the  generalized reciprocity~\cite{Beccaria:2009ny} present in the ${\cal N}=6$ superconformal Chern-Simons theory in three dimensions~\cite{Aharony:2008ug}.

We must stress that reciprocity \emph{is not} a rigorous prediction, in that it is still missing a first-principles derivation. Instead, it is  based on sound physical arguments and always needs to be verified, both at weak and at strong coupling.  However, its persistent validity is an intriguing empirical observation which can be at the moment qualified as a kind of hidden  symmetry of the integrable structures underlying the AdS/CFT system. Furthermore,  its  powerful predictive power  on the spectrum of the theories has been already successfully employed to formulate a five-loop analytic formula for the anomalous dimension of twist-three operators~\cite{BFLZ}, which has been confirmed by a purely field-theoretical calculation~\cite{Fiamberti:2009jw}~\footnote{With a similar reciprocity-based Ansatz a five-loop formula for the twist-two anomalous dimension was worked out in~\cite{5loopstwist2}.}.

\bigskip

The plan of this Review is the following. In Section 2 we recall the original Gribov-Lipatov formulation of the reciprocity property in QCD and sketch a modern reinterpretation of it as in~\cite{dok1,dok2} and~\cite{bk}. In Section 3 we present its generalized definition to the supersymmetric case  of $\mathcal{N}=4$ SYM theory. In Section 4.1 and 4.2, after a short introduction on the outcomes of integrability-based techniques, we collect  the information on the relevant multi-loop results for the anomalous dimensions of quasipartonic operators at weak coupling. We proceed then in Section 4.3 illustrating with specific examples how reciprocity has been checked on those anomalous dimensions, explaining then in Section 4.4 the way reciprocity can be used to produce new higher order formulas. Section 4.5 summarizes the weak coupling analysis. In Section 5 we present the strong coupling analysis of reciprocity, based on the perturbative (in the sigma model loop expansion) energies of  folded and spiky string solutions in $\ads$. 
The final Section 6 is devoted to a short list of open problems related to the subject of this Review. Three Appendices follow, in which we recall the basic properties of harmonic sums (Appendix \ref{app:harmonicsums}) and briefly illustrate the checks of reciprocity in the first commuting charges of the $\mathfrak{sl} (2)$ sector (Appendix \ref{app:higher}) as well as the generalized reciprocity of the so-called ABJM~\cite{Aharony:2008ug} theory (Appendix \ref{app:abjm}).

\section{Generalized Gribov-Lipatov reciprocity  in QCD}
\label{sec:reci}
 
 The anomalous dimensions $\gamma(S)$ of the twist-two  operators  with spin $S$ emerging in the  QCD analysis of deep inelastic scattering (DIS)~\cite{Altarelli:1981ax,Martin:2008cn} are expected to contain important information encoded in their dependence on $S$.
Connecting the total spin $S$ to its dual  in Mellin space,  the Bjorken variable $x$, two opposite regimes emerge in a natural way. 
The first is small $x\to 0$ and is captured by the BFKL equation. It can be analyzed by considering 
the Regge poles of $\gamma(S)$ analytically continued to negative (unphysical) values of the spin. 
 
Here, we shall be interested in the properties of the second {\em quasi-elastic} regime which is $x\to 1$, \emph{i.e.} large $S$.
From the large $S$ behavior of the known three loops twist-two QCD results
as well as from general results valid at higher twist~\cite{Belitsky:2003ys} the following general features can be inferred.
 The leading large $S$ behavior of the anomalous dimensions $\gamma(S)$ is logarithmic 
\be\label{cusp}
\gamma(S) = f(\lambda)\,\log\,S + {\cal O}(S^0),\qquad S\to\infty,
\ee
where $f(\lambda)$ is a universal function of the coupling related to soft gluon emission~\cite{Korchemsky:1992xv,Belitsky:2003ys,Belitsky:2006en}.
It appears as a cusp anomalous dimension governing the renormalization of a light-cone Wilson loop describing soft-emission processes as quasi-classical charge motion.
About the subleading $\sim \log^p\,S/S^q$ corrections, they
are found to obey special relations first investigated by Moch, Vermaseren and Mogt in~\cite{Moch} (see also, at two loops,~\cite{Curci:1980uw}) and known as MVV relations. 
Roughly speaking, they  predict the three-loop $1/S$ contributions in terms  of the $S^0$ two-loop ones.
The  MVV relations have received a relatively recent intriguing explanation in terms of  a non-trivial generalization of the one-loop Gribov-Lipatov  reciprocity~\cite{Gribov:1972rt} which is the subject of the next sections.

\subsection{Old Gribov-Lipatov reciprocity: a review}

In the QCD context the idea of \emph{reciprocity} arises from the attempt to symmetrically treat deep inelastic scattering (DIS) and the its crossed version, {\em i.e.} $e^+ e^-$ annihilation into hadrons.
 In DIS  the non-perturbative information is contained in the  space-like (S) splitting functions $P_S(x)$, governed by the Dokshitzer-Gribov-Lipatov-Altarelli-Parisi 
evolution~\cite{Gribov:1972rt,Altarelli:1977zs,Dokshitzer:1977sg} and  related to the anomalous dimensions $\g_S(S)$~\cite{Brock:1993sz}  of the twist-two operators via a Mellin transform. 
Instead,  the crossed process involves the non perturbative fragmentation functions, 
whose scale evolution is related  to the  time-like splitting functions $P_T(x)$; in this case the Mellin transform defines a time-like anomalous dimensions $\gamma_T(S)$.

The two types of  splitting functions were related by analytic continuation 
through the singular point~$x=1$  in the relation
worked out by Drell, Levy and Yan~\cite{DLY}
\begin{equation}
\mbox{Drell-Levy-Yan}:\qquad P_T(x) =
-\frac{1}{x}\,P_S\left(\frac{1}{x}\right).
\end{equation}
A second relation has been proposed by Gribov and
Lipatov~\cite{Gribov:1972rt}, stating an identical parton evolution for the
two processes
\begin{equation}
\mbox{Gribov-Lipatov}:\qquad P_T(x) = P_S(x) \equiv P(x).
\end{equation}
Combining the two relations above one can deduce a ``reciprocity property'' of the common function ${\cal P}(x)$
\begin{equation}\label{GLR}
\mbox{Gribov-Lipatov reciprocity}:\qquad P(x) = -x\,P\left(\frac{1}{x}\right).
\end{equation}
In Mellin space
\begin{equation}
 \cP(S) = \int_0^1\frac{dx}{x}\,x^S\,P(x) \equiv \Mellin{P(x)},
\end{equation}
 it can be shown~\cite{bk,Beccaria:2008fi} that this means (in the sense of asymptotic expansions at large $S$) 
\be
P(S) = f(C^2), \qquad C^2 = S\,(S+1),\qquad S\to\infty.
\ee
Gribov-Lipatov reciprocity holds at one-loop, but fails at two loops~\cite{Curci:1980uw,Furmanski:1980cm}.
The explicit violation can be written as
\be
\label{eq:violation}
\frac{1}{2}\left[P_{T, qq}^{(2)}(x)-P_{S, qq}^{(2)}\right] = \int_0^1\frac{dz}{z}\left\{P_{qq}^{(1)}\left(\frac{x}{z}\right)\right\}_+\,P_{qq}^{(1)}(z)\,\log\,z.
\ee

\subsection{Reciprocity respecting evolution equations}

The evolution equations for the parton distributions or fragmentation functions $D_\sigma(x, Q^2)$ ($\sigma = S, T$) take the 
standard convolution form 
\be
\partial_\tau\,D_\sigma(x, Q^2) = \int_0^1\frac{dz}{z}\,P_\sigma(z, \alpha_s(Q^2))\,D_\sigma\left(\frac{x}{z}, Q^2\right),
\ee
where $P_\sigma$ are the space or time-like splitting functions, $\alpha_s(Q^2)$ is the QCD running coupling constant and $\tau = \log\,Q^2$.
 Mellin transforming, this reads
\be
\partial_\tau\,D_\sigma(S, Q^2) = -\frac{1}{2}\gamma_\sigma(S, \alpha_s(Q^2))\, D_\sigma(S, Q^2),
\ee
where
\be
D_\sigma(S, Q^2) = \int_0^1\frac{dx}{x}\,x^S\,D_\sigma(x, Q^2),\qquad
\gamma_\sigma(S, Q^2) = -\frac{1}{2}\int_0^1\frac{dx}{x}\,x^S\,P_\sigma(x, \alpha_s(Q^2)).
\ee
Based on several deep physical ideas, it has been proposed to rewrite the evolution equation in a way that aims at treating the DIS and $e^+e^-$ channels
more symmetrically, in the spirit of Gribov-Lipatov reciprocity~\cite{Dokshitzer:1995ev,Dokshitzer:2005bf}.
The reciprocity respecting evolution equations  take the form 
\be
\label{eq:REE}
\partial_\tau\,D_\sigma(x, Q^2) = \int_0^1\frac{dz}{z}\,{\cal P}(z)\,D_\sigma\left(\frac{x}{z}, z^\sigma\,Q^2\right),
\ee
where $\sigma=-1, 1$ for the space-like and time-like channels respectively. 
The crucial point is that the evolution kernel ${\cal P}(z)$ is the same in both channels. 
As an immediate check, one recovers for the non-singlet quark evolution the Curci-Furmansky-Petronzio relation~\refeq{violation}.
Other features related to the Low, Burnett, Kroll theorems~\cite{Low:1958sn} (LBK) as well as to 
the {\em inheritance} idea are further discussed in~\cite{Dokshitzer:2005bf}.
A successful three loop check using the $\gamma_T$ evaluated by Drell-Levy-Yan analytic continuation is 
described in~\cite{Mitov:2006ic} for the non-singlet QCD anomalous dimensions.

\subsection{Moch-Vermaseren-Moch relations and reciprocity of the kernel ${\cal P}$}

The previous formulation of reciprocity is in $x$-space, but has important consequences in the large spin expansion of the anomalous dimensions. This point of view is adopted in Basso and Korchemsky~\cite{bk} who 
propose a very simple way of testing  \refeq{REE}.

Neglecting effects due to the running couplings~\footnote{We are going to discuss $N=4$ SYM which is 
ultraviolet finite.}, one immediately derives from \refeq{REE} the non-linear relation 
(after a rescaling of ${\cal P}$)
\be\label{nonlinearQCD}
\gamma_\sigma(S) = {\cal P}\left(S-\frac{1}{2}\,\sigma\,\gamma_\sigma(S)\right).
\ee
In the spirit of the derivation of the reciprocity respecting evolution equation \refeq{REE} it is natural to
guess that the Mellin transform of the kernel ${\cal P}$ in (\ref{nonlinearQCD}) obeys the Gribov-Lipatov reciprocity relation (\ref{GLR}).

As an immediate corollary, the following general parametrization of the large $S$ expansion of $\gamma_\sigma$ (we define $\overline S = S\,e^{\gamma_{\rm E}}$
and $A = f(\l)$)
\be\label{genexp}
\gamma_\sigma(S) = A\,\log\,\overline{S} + B + C_\sigma \,\frac{\log\,\overline{S}}{S} + \left(D_\sigma+\frac{1}{2}\,A\right)\frac{1}{S} + \cdots, 
\ee
must satisfy the constraints
\be
\label{eq:mvv1}
C_\sigma = -\frac{1}{2}\,\sigma\,A^2, \qquad 
D_\sigma = -\frac{1}{2}\,\sigma\,A\,B,
\ee
which are highly non-trivial since $A, B, C$ and $D$ are functions of the gauge coupling.
The first relation in (\ref{eq:mvv1}) is indeed verified at three loops by the explicit evaluation of $\gamma_\sigma$,  being part of the above-mentioned MVV relations.
Most importantly, as discussed in~\cite{bk}, the second (subleading) relation requires, in QCD, a correction in the relation (\ref{nonlinearQCD})  that is related to the non-zero value of the $\beta$ function.
For twist-two operators in the finite ${\cal N}=4$ SYM theory, it is correct as it stands.

Thus, the two MVV relations in \refeq{mvv1} strongly suggest that, when formulated for the Mellin transform of the kernel $\cP$ defined in (\ref{nonlinearQCD}),  the reciprocity relation (\ref{GLR}) holds.
In $S$-space, it is equivalent to the claim that ${\cal P}(S)$ has a large $S$ expansion in {\em   integer powers of } $C^2$ of the form 
\be
\label{eq:firstreciprocity}
{\cal P}(S) = \sum_n\frac{a_n(\log\,C)}{C^{2n}},
\ee
where $C^2 = S\,(S+1)$, and $a_n$ are polynomials which can be computed in perturbation theory as series in $\alpha_s$.
The expansion (\ref{eq:firstreciprocity}) can be read as a parity invariance under $S\to -S-1$, although this must be considered only as an analytic continuation around $S=\infty$ and not at any $S$
in strict sense  because of the Regge poles at negative $S$.

The property (\ref{eq:firstreciprocity}), or its equivalent form (\ref{GLR}), has indeed been checked at three loops in~\cite{bk}  for several classes of twist-two operators in QCD.
It generates an infinite set of MVV-like relations for all the subleading terms in the large $S$ expansion of the anomalous dimensions. The previous relations \refeq{mvv1} are just the first cases.

\section{Generalized reciprocity in ${\cal N}=4$ SYM}
\label{sec:n4}

Reciprocity has been discussed in QCD, a theory which shares the gluon sector with ${\cal N}=4$ SYM. This suggests to explore its validity in the latter, highly symmetric theory where one can exploit integrability to compute multi-loop anomalous dimensions.
 
 Since the leading order evolution kernel of  ${\cal N}$ = 4 SYM theory is purely classical in the LBK  sense~\cite{Dokshitzer:2009wd}, there is hope to derive one day  
 all-loop expressions for the anomalous dimensions  of the operators of the  theory within a \emph{simple} description, \emph{i.e.} in which higher order terms are dynamically 
 inherited from the first loop. QCD would greatly benefit from such a result, and in general from investigations in which  ${\cal N} = 4$ SYM is studied with the aim of putting 
 under full theoretical control the dominant part of the perturbative QCD gluon dynamics.

\medskip

The conformal invariance of the $\mathcal{N}=4$ SYM theory allows one to extend the results of the previous section to the anomalous dimensions
of the so-called \emph{quasipartonic operators} of arbitrary twist $J$. The definition of the quasipartonic operators~\cite{quasipart} goes back to the conformal limit
of the QCD, and is in fact unrelated to the presence of supersymmetry.

In the conformal limit, the light-cone ray is left invariant by a $SL(2,\mathbb{R})$ collinear subgroup of the conformal group, generated by translations and dilatations along the ray, and rotations in the $x^{0}\pm x^{1}$ plane~\cite{bra}. 
In light cone gauge, one can identify preferred components ($SL(2,\mathbb{R})$ primary fields) of the elementary scalars (in supersymmetric theories), Weyl fermions and field strength   with minimal collinear twist~\footnote{$SL(2,\mathbb{R})$ primary fields $\Phi$ have definite scaling dimension $d$ and collinear spin $c$ defined by 
\be
D\,\Phi=d\,\Phi,~~~~~~~~\Sigma_{+-}\,\Phi=c\,\Phi,
\ee
where $D$ and $\Sigma_{\mu\nu}$ are the dilatation and Lorentz spin generators. The collinear twist  (collinear dimension minus collinear spin) is minimal for $t=d-c=1$.}.
Composite operators built with 
this set and an arbitrary number of covariant derivatives correspond to
\emph{physical} degrees of freedom, as it is clear in light-cone gauge, and are called quasipartonic operators.

We shall then write a general quasipartonic single trace gauge invariant operator as
\begin{equation}
\label{eq:O}
\hat{\mathcal{O}}(z_1, \dots, z_J) = {\rm Tr}\big\{X(z_1\,n)\cdots X(z_J\,n)\big\},
\end{equation}
where $z\,n^\mu$ is the light-like ray and $X$ can be a (suitable) ${\cal
  N}=4$ scalar field $\varphi$, gaugino component $\lambda$, or 
holomorphic combination $A$ of the physical gauge field $A^\mu_\perp$~\cite{bra}.  The number of the constituent fields $J$ is the twist (classical dimension minus spin) of the operator. 

At one-loop these operators have simple transformation properties with respect to the
collinear group, they transform as 
$[\ell]^{\otimes L}$ where $[\ell]$ is the 
infinite-dimensional $\mathfrak{sl}(2)$ representation with \emph{conformal spin} 
respectively~\cite{bra}
\begin{equation}
\label{eq:confspin}
\ell(\varphi) = \frac 1 2,\qquad
\ell(\lambda) = 1,\qquad
\ell(A) = \frac 3 2.
\end{equation}

\medskip

 A suitable generalization of the analysis of reciprocity in Refs.~\cite{Dokshitzer:2005bf,bk} to the case of $\mathcal{N}=4$ SYM  assumes that $\g(S)$ obeys at all orders the non-linear equation~
 \footnote{Since by $\g(S)$ one means the anomalous dimension  of a gauge invariant operator in $\mathcal{N}=4$ SYM theory,  it is quite natural to adopt for such generalization the case of $\sigma=-1$ in the nonlinear  QCD relation (\ref{nonlinearQCD}), corresponding to the space-like case. In fact, the QCD time-like anomalous dimensions are \emph{not} related to composite local gauge operators, due to the general fact that fragmentation functions do not admit the operator product expansion~\cite{bk}.}
\be\label{Nonlinear}
\gamma(S) = {\cal
  P}\left(S+\frac{1}{2}\, \gamma(S)\right),
\ee
and the  reciprocity relation takes the form
\be
\label{Parity}
{\cal P}(S) = \sum_n\frac{a_n(\log\,C)}{C^{2n}},
\ee
where $a_n(\log\,C)$ are suitable polynomials, $C$ is obtained by replacing $S(S+1)$ with the  Casimir of the 
collinear conformal subgroup $SL(2, \mathbb{R})\subset SO(4,2)$
\be
\label{Casimir}
C^2 =s(s-1)\equiv (S+J\,\ell-1)\,(S+J\,\ell).
\ee
Here, $s=\frac{S+\Delta_0}{2}=S+J\,\ell$ is the ''bare'' conformal spin $s$ of the operator (with $\Delta_0$ being the canonical dimension of the operator) defined in terms of the conformal spin $\ell$   of the fields (\ref{eq:confspin}) out of which the operator is built. The constraint (\ref{Parity}) is simply a parity invariance under (large) $C\to-C$.

This generalization is related to the proposal by~\cite{bk} of tracing back the origin of the nonlinear relation (\ref{Nonlinear}) to the conformal symmetry of the theory (for the same reason, and as mentioned above, in gauge theories with nonvanishing beta-function, like QCD, the anomalous dimensions receive conformal symmetry breaking contributions). 
Quasipartonic  operators can be in fact classified according to
  representations of the collinear $SL(2, \mathbb{R})$ subgroup of the
   $SO(2,4)$ conformal  group  which are labeled by the
conformal spin  of the operator~ \ci{bra}, whose general definition $s=\frac{S+\Delta}{2}$ involves in fact the scaling dimension of the operator. Since this get renormalized, receiving anomalous contribution $\g$ at higher orders, one may argue that the anomalous dimension itself  should be a function
     of  $S$ only through its dependence on the ``renormalized'' conformal spin redefined in terms of    $\Delta = S + J + \g(S,J)$.    This then leads to the nonlinear 
functional relation     for $\g$\foot{    The relation between the notation used in~\cite{bk} and ours  is:  $N\to S$, $L\to J$,  $J\to C$ and $j\to  s$.}
   \be \g(S,J) = f(s; J)\equiv f\,\Big( S + \ha J + \frac{1}{2} \g(S,J); J\Big) \ .
     \label{ko}
      \ee
   Suppressing the dependence on $J$ in $\g$ and $f$ one may write such functional relation simply as (\ref{Nonlinear}).

One can notice that without further information eq. (\ref{Nonlinear}) is nothing more than a change of variable, since, at least in perturbation theory, it is always possible to compute the
 function $\PP$ in terms of  the anomalous dimension $\g(S, J)$.   The non-trivial information is in fact contained in the parity invariance (\ref{Parity}), from which an infinite set of constraints can be derived between subleading coefficients in a general large spin expansion of the anomalous dimension, exactly as it happens in eqs. (\ref{genexp}) and (\ref{eq:mvv1}) above.

\subsection{Strong form of reciprocity from the simplicity of ${\cal P}$}
We conclude this section with some interesting observation about  the large spin expansion of the function $\cP$. Its leading logarithmic behavior, as follows from the structure of (\ref{Nonlinear}), coincides with the leading behavior of $\gamma$ in (\ref{cusp}), where the coupling dependent scaling function $f (\lambda)$ (cusp anomaly) is expected to be universal in {\em both twist and flavour}~\cite{Belitsky:2003ys,Eden:2006rx}.  Concerning the subleading terms, 
as remarked in ~\cite{bk,dok2},  the function $\cP(S)$ obeys \emph{up to  three loops} 
a  powerful additional {\em simplicity} constraint, in that it does not contain  logarithmically enhanced 
terms $\sim \log^n(S)/S^m$ with $n\ge m$. This immediately implies that the leading logarithmic functional relation 
\be\label{resum}
 \gamma(S)=f (\lambda)\,\log\left(\textstyle{S+\frac{1}{2}f (\lambda)\log S+} ...\right)+...
 \ee
predicts correctly the maximal logarithmic terms $\log^m{S}/S^m$ 
\be\label{leadinglogs}
\gamma(S)\sim f\,\log S+\frac{f ^2}{2}\,\frac{\log S}{S}-\frac{f ^3}{8}\,\frac{\log^2 S}{S^2}+...
\ee
whose coefficients are simply proportional to $f ^{m+1}$~\cite{bdm,bkp,BFTT}.

Notice that the fact that the cusp anomaly is  known at all orders in the coupling via the results of~\cite{Beisert:2006ez,bassocusp} would in principle imply 
(\emph{under the ``simplicity'' assumption for  $\cP$}) a proper \emph{prediction} for such maximal logarithmic terms at \emph{all orders} in the coupling constant, and in particular for those appearing in the large spin 
expansion of the energies of certain semiclassical string configurations (dual to the operators of interest). 
As we will report in the sections dedicated to the strong coupling checks of reciprocity, such ``inheritance'' has indeed been  checked in~\cite{BFTT}  up to one loop in the sigma model semiclassical expansion, as well as 
in~\cite{Ishizeki:2008tx} at the classical level. An independent strong coupling confirmation 
of (\ref{leadinglogs}) up to order $1/S$   has recently been given for twist-two operators in~\cite{Freyhult:2009my}. 

However, the asymptotic part of the four-loop anomalous dimension for twist-two operators and of the  five-loop anomalous dimension for twist-three operators reveal an exception to this ``rule'', being the term $\log^2 S/S^2$  not given only in terms of the cusp anomaly~\footnote{Interestingly enough, the large spin expansion of the wrapping contribution of~\cite{Bajnok:2008qj} and of~\cite{BFLZ}, which correctly does not change the leading asymptotic behavior (cusp anomaly), first contributes at the same order, but \emph{not} in such a way that the total $\log^2 S/S^2$ coefficient results in $\big(\textstyle{-\frac{f^3}{8}}\big)$ as required from (\ref{leadinglogs}).
}.
This seem to indicate that, at least for twist-two and twist-three operators in the $\mathfrak{sl}(2)$ sector and at critical wrapping order,  the $\cP$-function ceases to be ``simple'' in the meaning of ~\cite{dok2}, thus preventing the tower of subleading logarithmic singularities $\log^m S/S^m$ to be simply inherited from the cusp anomaly. In order to clarify how the observed difference in the simplicity of the $\cP$ at weak and strong coupling works, further orders  in the semiclassical sigma model expansion would be needed.

\section{Reciprocity tests at weak coupling in ${\cal N}=4$ SYM}

Given our interest in testing reciprocity in ${\cal N}=4$ SYM, the next step is to exploit 
integrability in this theory to achieve closed form for $\gamma(S)$ of specific classes of operators
at many loops.

\subsection{Multi-loop calculation of  anomalous dimensions via integrability}
  \label{sec:bethe}

The calculation of the anomalous dimensions in the planar limit of $\mathcal{N}=4$ SYM theory
is in fact dramatically simplified by its integrability properties. The gauge theory composite operators can be 
mapped to states of a $PSU(2,2|4)$ invariant integrable spin chain, which for quasipartonic operators coincides at one loop with the $\mbox{XXX}_{-\ell}$ 
chain~\cite{Braun:1998id}. The energy of the spin chain 
is the image of the gauge theory dilatation operator. Thus, the calculation of the coupling dependent
energy levels of the spin chain provides the multi-loop anomalous dimension of specific gauge theory 
composite operators.

We can illustrate the general strategy with a specific example which will be relevant in the 
following discussion. We consider the subsector $\mathfrak{sl}(2)\subset\mathfrak{psu}(2,2|4)$
which is perturbatively closed at all orders under renormalization. This sector contains composite operators
which can be written schematically as ${\cal O}_{J, S} = \varphi^{J-1}{\cal D}^{S}\varphi$, where $\varphi$ is a scalar field and $\cal D$ a certain projected covariant derivative.

The integrable structure of the spin chain,  the conformal spin (\ref{eq:confspin}) being here $\ell=\frac{1}{2}$, leads to the following Bethe equations at one-loop
\begin{equation}
\left(\frac{u_k+\frac{i}{2}}{u_k-\frac{i}{2}}\right)^J = \prod_{j=1, j\neq
  k}^S\frac{u_k-u_j-i}{u_k-u_j+i},
\end{equation}
where $u_{i}$ are the Bethe roots, in terms of which is written the one-loop anomalous dimension 
\begin{equation}
\gamma_1 = \sum_{k=1}^S\frac{1}{u_k^2+\frac{1}{4}}.
\end{equation}
The same equations can be conveniently reformulated in the language of the Baxter operator. In this 
simple context, one considers the polynomial 
$Q(u) = \prod_{i=1}^{S}(u-u_{i})$ which obeys the equation
\begin{equation}
\big(u+\frac{i}{2}\big)^J\,Q(u+i) + \big(u-\frac{i}{2}\big)^J\,Q(u-i) = t(u)\,Q(u),
\end{equation}
where $t(u)$ is the transfer matrix of the integrable chain.
In terms of $Q(u)$, the one-loop anomalous dimension reads
\begin{equation}
\gamma_1 = i\,\left. (\log Q(u))' \vphantom{\frac{1}{2}}
\right|_{u=-\frac{i}{2}}^{u=+\frac{i}{2}}.
\end{equation}
In the simplest case of twist $J=2$ the transfer matrix is a  second order
polynomial $t(u)=2u^2-(S^2+S+1/2)$, and the solution is easily identified with
the Hahn function $Q(u)={}_3 F_{2}(-S, S+1, 1/2-i u;1,1;1)$. Thus, the anomalous
dimension at the 1-loop order, or $g^2=\frac{\gym^2 N}{16\,\pi^2}$, is 
\begin{equation}
\gamma(S)= 2g^2 \left( Q'(i/2) - Q'(-i/2) \right)=8\, g^2 \,S_1(S),\qquad S_{1}(S) = \sum_{n=1}^{S}
\frac{1}{n},
\end{equation} 
This construction can be extended to all loops both in terms of  Bethe Ansatz equations~\cite{Beisert:2005fw}
as well as with the Baxter formalism ~\cite{Bel06,Bel09,Beccaria:2009rw,KotRejZie08}.

In principle, the Baxter method is superior to the other, since it provides an analytical expression 
to the anomalous dimension as a function of the number  $S$ of Bethe roots. 
Nevertheless, this approach has not been pursued in full details for higher rank subsectors of the 
theory and a practical alternative is the  {\em  maximal
transcendentality principle} ~\cite{MTP}.

This QCD-inspired idea~\footnote{Inspired by the structure of the two loop anomalous dimension of $\mathcal{N}=4$ twist two  operators in the $\mathfrak{sl}(2)$ sector, it has been proposed~\cite{MTP}  that the three-loop answer could be extracted by simply picking up the ``most transcendental terms''  from the three-loop non-singlet QCD anomalous dimension derived in~\cite{Moch}. The conjectured three loop formula has been then independently confirmed in the framework of the Bethe ansatz equations ~\cite{Staudacher:2004tk} as well as within a space-time approach~\cite{Bern:2005iz}.} predicts that at each order $n$ the solution can be entirely expressed in terms of certain combinations of generalized harmonic sums 
of order $2n - 1$ \emph{or} in terms of products of harmonic sums  $S_{\mathbf{a}}$ and 
zeta functions $\zeta(b_{i})$ in such a way that the sum of their transcendentalities $|\mathbf{a}|$ and $b_{i}$ (see Appendix A for definitions) is again equal to $2n -1$. One can then use the 
maximal transcendentality principle to write
the anomalous dimensions as combination  of harmonic sum of fixed order with
coefficient to be determined. The {\em rational} coefficients can be
then computed by fitting numerically with high precision  the perturbative expansions of the
Bethe equation at fixed  $S$.\\

A crucial point   is that the
derivation of the Bethe equations, or equivalently of the Baxter equation, is based on the assumption that the 
length of the composite operator, {\em i.e.} the spin chain length, is sufficiently large to avoid 
finite size effects related to interactions which \emph{wrap} around the chain.  The additional {\em wrapping contributions}
which occur for short chains were for the first time correctly evaluated in~\cite{Bajnok:2008bm}  via a clever generalization   of the L\"uscher formulas \cite{Luscher:1985dn} previously proposed for the $\ads$ sigma model in~\cite{Janikprev}.
 Such finite size effects are the object of recent investigations exploiting thermodynamical Bethe Ansatz methods and relying on the AdS/CFT duality with the superstring dynamics on $\ads$~\cite{Ysystem, TBA}. The general statement is then that  the full anomalous dimension must be written as
\be\label{fullg}
\gamma(g) = \gamma^{\rm ABA}(g) + \gamma^{\rm wrapping}(g), 
\ee
where $\gamma^{\rm ABA}(g)$ is captured by the  asymptotic Bethe Ansatz  equations of~\cite{Beisert:2005fw}  and 
$\gamma^{\rm wrapping}(g)$ is the wrapping contribution that can be evaluated with the tools mentioned above.

From the point of view of this Review, it is expected that reciprocity holds for the full anomalous dimension (\ref{fullg}), since the above splitting has a more technical than physical nature.
In all the explored examples to be discussed in the next section, reciprocity holds in fact for both the asymptotic and the wrapping part. It is however remarkable that this happens \emph{separately} for the individual contributions.

\subsection{Applications to quasipartonic composite operators}
  \label{sec:examples}

We collect here the information on the relevant multi-loop results for the anomalous dimensions 
 of a class quasipartonic operators in $\mathcal{N}=4$
SYM. The discussion about the reciprocity properties of these results will follow in the next section.

As mentioned in Section 4.1, the emergence of integrability in the planar limit allows one to construct (at
least at the one-loop level) a dictionary of correspondences between quasipartonic operators and generalized spin chains. In the spin-chain language
quasipartonic operators correspond to fixed length states and the anomalous dimensions are
 the hamiltonian eigenvalues of the relevant $\mbox{XXX}_{-\ell}$
chain~\cite{Braun:1998id}.
At the one loop level these sets are closed under perturbative
renormalization, while at higher loops only the operators built out of scalar fields and gauginos continue to scale autonomously. In fact, the ${\cal N}=4$ $\mathfrak{sl}(2)$ subsector  is closed at all orders, and even though operator with gauginos span the $\mathfrak{sl}(2|1)$
subsector where there is mixing between scalars and fermions, this is not true in the quasipartonic set of operators built out of   suitably projected  components of gaugino fields~\cite{Belitsky:2007zp}.
Finally, in the case of {\em gauge operators}~\footnote{The name stems from the one-loop description of 
a class of scaling operators. Beyond one-loop, additional fields mix.}~\cite{Belitsky:2008wj},   mixing effects start immediately beyond one-loop (see the discussion in~\cite{Beccaria:2007pb}).\\

\noindent\underline {\em  a. Scalar Operators}\\

The most studied and simplest sector is the  $\mathfrak{sl}(2)$
subsector of the theory,  whose representative operators ${\cal O}_{J, S} = \varphi^{J-1}{\cal D}^{S}\varphi$, built out of scalar fields $\varphi$ and covariant derivatives acting on them, were introduced in Section 4.1. In the chain language each covariant derivative  is thought as an ``excitation'' of the vacuum state ${\rm Tr}\, \varphi^J$. The number of these excitations $S = \sum n_i$, the total spin, is not limited,  being the  $-\frac{1}{2}$ representation of $\mathfrak{sl}(2)$ infinite-dimensional.

The relevance of this bosonic subsector is due to the fact that, in the important case of twist-two operators, it is exhaustive of the whole theory. All twist-two operators fall in fact in a single
supermultiplet~\cite{Belitsky:2003sh,Belitsky:2005gr,Beisert:2002tn} and
their anomalous dimension is expressed in terms of a universal function $\gamma_{\rm univ}$  with shifted arguments
\be\label{twist2universal}
\g^\varphi_{J=2}(S)=\g_{{\rm univ}}(S)~,~~\g^\psi_{J=2}(S)=\g_{{\rm univ}}(S+1)~,~~\g^A_{
J=2}(S)=\g_{{\rm univ}}(S+2).
\ee

For the twist-two anomalous dimensions, closed expressions at two loops  are known from explicit field-theory calculations~\cite{Kotikov:2003fb} and at three-loops from
a  conjecture  inspired from the  maximum transcendentality principle~\cite{MTP} applied to  the QCD splitting functions~\cite{Moch}. Up to three loops, the same formulas can  also be  computed by the asymptotic Bethe ansatz~\cite{Staudacher:2004tk} for fixed values
of $S$. It is only recently that the three loop conjecture has been proved via the Baxter approach method~\cite{Kotikov:2008pv}. In~\cite{Kotikov:2007cy} and~\cite{Bajnok:2008qj} the ABA and wrapping part for the four-loop anomalous 
dimensions for twist two scalar operators in the $\mathfrak{sl}(2)$ have been
computed, with the techniques explained in the previous section. This result has been confirmed by a  field-theoretical calculation~\cite{Fiamberti:2007rj,Veliz}. With similar ABA techniques and in absence of wrapping corrections, closed (in $S$) expressions for the anomalous dimensions of twist-three operators were derived in~\cite{Kotikov:2007cy} and~\cite{Beccaria:2007cn}. 

Exploiting an Ansatz based on reciprocity (see next section), a five-loop formula for the anomalous dimensions was proposed in~\cite{BFLZ} 
 for the twist-three operators and, in a similar fashion, in~\cite{5loopstwist2} for the case of twist-two.
 While in the first  $J=3$ case the formula involves a leading order (generalized) L\"uscher correction, in the case of $J=2$ a non-trivial  next-to-leading order wrapping contribution (together with a modification of the quantization condition) comes into play. This is  due to the general fact that, in the $\mathfrak{sl}(2)$ sector, for twist $J$  
operators the wrapping effect starts at order $g^{2J+4}$,  delayed by superconformal invariance.
The twist-three five-loop formula has been later confirmed by a purely field-theoretical calculation~\cite{Fiamberti:2009jw}, while the correctness of the recent five-loop twist-two proposal is strongly supported by the fact that it respects the correct weak-coupling constraints deriving from a BFKL analysis and double-logarithmic behavior.

The same techniques used for the anomalous dimensions work in the case of the higher conserved charges of the chain model~\cite{Arutyunov:2003rg}, something discussed so far only for the first few charges in the scalar sector~\cite{Beccaria:2009yt} and reviewed in Appendix B.\\

\noindent\underline {\em  b. Fermion operators}\\

These operators are built out of   helicity~$+\frac{1}{2}$ component of the gaugino fields $\lambda_\alpha$, and covariant derivatives acting on them, defined in~\cite{Belitsky:2005bu}, where twist-three representatives have been studied at two loops in ${\cal N}=1, 2, 4$ SYM by direct computation of  the dilatation operator. 
The high level of symmetry of the $\mathcal{N}=4$ theory results in a number
of degeneracies in the spectrum of anomalous dimensions, with unexpected
relations between composite operators of different twist~\cite{Beisert:2005fw}.
The Bethe Ansatz reflects of course such remarkable structural 
properties related to supersymmetry.
 
An excellent example of this fact is precisely the case of twist three operators
 built out of  gauginos whose anomalous dimension was first proved  in~\cite{Beccaria:2007vh}  to be related to the  ``universal'' twist two anomalous dimension (\ref{twist2universal}) as
\be\label{gauginoshift}
\gamma^\psi_{J=3}(S) = \gamma^\varphi_{J=2}(S+2).
\ee
This statement has been rigorously proved at three loops and attributed to a hidden
$\mathfrak{psu}(1|1)$ invariance of the $\mathfrak{su}(2|1)$ subsector of the theory.\\

\noindent\underline {\em  c. Gauge operators}\\

These quasipartonic operators have as constituents gauge
fields $A$ on which an arbitrary number of covariant derivatives act, where $A$ stands for  the holomorphic combination of the physical gauge degrees of
freedom $A^\mu_\perp$ (suitable projected components of the field strength) defined in~\cite{bra}.
Twist-three gauge operators were considered in~\cite{Beccaria:2007pb} at three loops, and in~\cite{Beccaria:2008fi} at four loops and without wrapping effects. 
 
At one-loop, this sector is described by a non-compact $XXX_{-3/2}$  spin chain with $J$ sites, and the anomalous dimension is known as an exact solution of the Baxter equation. Beyond this order, no simple spin-chain correspondence exist and mixing effects come into play. In order to find a closed formula for the anomalous dimension, one can then hope to make use of the full $\mathfrak{psu}(2,2|4)$ Bethe equations in which the quantum numbers belonging to the correct superconformal primary that describes this sector have to appear.  This can be done exploiting the superconformal properties of the (maximally symmetric) tensorial product of three singletons~\cite{Beisert:2004di}. As usual, using as an input the one-loop solution
\be
\gamma_{J=3}^A(S) =4\,S_1\left(\frac{S}{2}+1\right)-5+\frac{4}{S+4}.
\ee
one can solve numerically the Bethe equations order by order in perturbation theory and fit the coefficient in an appropriate Ansatz. However, in this case the latter cannot be inspired by the standard  maximum transcendentality principle, which is  violated already at one loop
as shown explicitly from the formula above. The latter  is fully consistent with the QCD analysis of maximal
helicity 3-gluon operators~\cite{Belitsky:1999bf}, where the dilatation operator
can be decomposed as an integrable piece ${\cal H}_0$ plus a perturbation and the lowest
eigenvalue is
\ba
\varepsilon = 4\,S_1\left(\frac{S}{2}+1\right)+\frac{4}{S+4}+4.
\ea
Inspired by a similar QCD calculation~\cite{Mertig:1995ny}, the following Ansatz can be made
which generalizes the one-loop result  at $k$ loops
\be
\label{Ansatzgauge}
\gamma_k(n) = \sum_{\tau=0}^{2\,k-1}\gamma^{(\tau)}(n), ~~~~~
\gamma^{(\tau)} (n) = \sum_{p+\ell = \tau}\frac{{\cal H}_{\tau,\ell}(n)}{(n+1)^p}, \qquad n = \frac{S}{2}+1,\nonumber
\ee
where ${\cal H}_{\tau,\ell}(n)$ is a combination of harmonic sums with homogeneous fixed transcendentality $\ell$. The terms with $p=0$ have maximum  transcendentality, all the others have subleading  transcendentality. Making use of this Ansatz and in the usual way, a three-loop~\cite{Beccaria:2007pb} and a four-loop formula~\cite{Beccaria:2008fi} were derived  for the anomalous dimension of these twist-three gauge operators.
 
 \subsection{Proof of reciprocity in closed form}

Reciprocity is checked on the function $\cP$ which is obtained inverting (\ref{Nonlinear}) as
\ba\label{LagrangeBuermann}
{\cal P}(S) = \sum_{k=1}^\infty \frac{1}{k!}\Big(-\frac{1}{2}\partial_S\Big)^{k-1}[\gamma(S)]^k = \gamma-\frac{1}{4}\,(\gamma^2)'+\frac{1}{24}\,(\gamma^3)''-\frac{1}{192}\,(\gamma^4)''' + \cdots .
\ea
inheriting thus the perturbative expansion of the anomalous dimension
\be
\gamma = \sum_{k=1}^\infty g^{2\,k}\,\gamma_k\quad\quad\quad {\cal P} = \sum_{k=1}^\infty g^{2\,k}\,{\cal P}_k
\ee
One way to operate  is checking directly the parity invariance (\ref{Parity}). One should perform the large $S$ expansion of (\ref{LagrangeBuermann}), rewrite it as a large $C$ expansion  inverting (\ref{Casimir}) and check the absence of odd inverse powers of $C$. Three-loop tests of reciprocity for QCD and for the universal twist-two supermultiplet in $\mathcal{N}=4$ SYM were discussed this way in~\cite{bk}, and it is also the procedure adopted up to now in the strong coupling analysis of reciprocity (see Section 5). At weak coupling, however, there is a much more elegant and powerful way to proceed. Considering that each term of the perturbative expansion of $\cP$ is a linear combination of products of harmonic sums, the idea is to find a new basis for the harmonic sums with definite properties under the (large-)$C$ parity $C\to-C$.

This has been done in~\cite{Beccaria:2009vt}, where the map $\omega_a$, $a\in \mathbb{N}$ has been introduced, which acts linearly on linear 
combinations of harmonic sums as follows~\footnote{We omit, in the following,  the dependence of the harmonic sums  on the spin $S$.}
\begin{equation}
\label{map}
\omega_a(S_{b, \mathbf{c}}) = S_{a, b, \mathbf{c}}-\frac{1}{2}\,S_{a\wedge b,
  \mathbf{c}},
\end{equation}
where, for $n, m\in \mathbb{Z} \backslash  \{0\}$, the wedge-product is
defined as
\begin{equation}
\label{wedge}
n\wedge m = \mbox{sign}(n)\,\mbox{sign}(m)\,(|n| + |m|).
\end{equation}
One can also consider a complementary map $\underline{\omega_a}$ acting in a similar way on complementary sums defined in appendix \ref{app:harmonicsums}.

Following~\cite{dok2,bdm}, the combinations of
(complementary) harmonic sums can be introduced~\footnote{A different basis for harmonic sums with well-defined reciprocity-respecting properties  has been recently proposed in~\cite{5loopstwist2}.}
\begin{eqnarray}
\begin{array}{lll}
\Omega_a & = & S_a, \\
\Omega_{a, \mathbf{b}} & = & \omega_a( \Omega_\mathbf{b}),
\end{array}
\qquad
\begin{array}{lll}
\underline{\Omega_a} & = & S_a = \underline{S_a}, \\
\underline{\Omega_{a, \mathbf{b}}} & = & \underline{\omega_a}(
\underline{\Omega_\mathbf{b}}).
\end{array}
\end{eqnarray}
for which the following two theorems hold~\cite{Beccaria:2009vt}.

\bigskip
\noindent
{\bf Theorem 1:}~\footnote{A special case of Theorem 1 appeared
  in~\cite{dok2}. A general proof of Theorem 1 in the restricted case 
$\mathbf{a} = (a_1, \dots, a_\ell)$ with {\em  positive} $a_i>0$ and  {\em 
  rightmost indices} $a_\ell\neq 1$ can be found in~\cite{bdm}.} 
{\em  The
subtracted complementary  combination
$\underline{\widehat\Omega_{\mathbf{a}}}$, $\mathbf{a} = (a_1, \dots, a_d)$
has definite parity 
$\cal P_\mathbf{a}$
under the (large-)$C$ transformation  $C\to -C$ and
\be
{\cal P}_\mathbf{a} = (-1)^{|a_1|+\cdots + |a_d|}\,(-1)^d\,\prod_{i=1}^d
\varepsilon_{a_i}.
\ee
}
\noindent
{\bf Theorem 2:}  {\em  The combination $\Omega_{\mathbf{a}}$, $\mathbf{a} =
  (a_1, \dots, a_d)$ with odd positive $a_i$ and 
even negative $a_i$ has positive parity ${\cal P}=1$.}
 

The strategy to prove the reciprocity property of the kernel $\cP$ is then the following.
At each perturbative order $\ell$ one  starts from the expression of the kernel 
$\cP_\ell$ written in the canonical basis, something that can always be done using the shuffle algebra  (\ref{eq:shuffle}), and isolate in this expression  
the sums  with maximum depth. Each of them, say $S_\mathbf{a}$,
appears uniquely as the maximum depth term in $\Omega_\mathbf{a}$.
One then subtracts all the $\Omega$'s required to cancel these terms, 
keeping track of this subtraction and repeating the procedure decreasing the depth by one. If one ends the algorithm with a zero remainder and the full subtraction is composed
by $\Omega$'s with the right parities (see Theorem 2), one can conclude that the
kernel $\cP$ is parity respecting at the investigated order.

For example, the four-loop wrapping contribution fro twist two anomalous dimension calculated in~\cite{Bajnok:2008qj} 
\ba\label{wrapping}
\gamma_4^{\rm wrapping}(S) &=& 256 \left(S_{-5}-S_5+2 S_{-2,-3}-2 S_{3,-2}+2 S_{4,1}-4 S_{-2,-2,1}\right) S_1^2+\nonumber\\
&& -640\, \zeta_5 \,S_1^2-512\, S_{-2}\, \zeta_3 \,S_1^2.
\ea
can be conveniently rewritten only in terms of allowed $\Omega$'s 
\be\label{P4wr}
\cP_4^{\rm wrapping} = -128\,\Omega_1^2\,(5\,\zeta_5+4\,\zeta_3\,\Omega_{-2}+8\,\Omega_{-2,-2,1}+4\,\Omega_{3, -2}).
\ee
This way reciprocity was proven at  four loops for the whole (ABA part included) anomalous dimension of twist-two operators. In a totally similar way,  four loop reciprocity tests have been performed for twist-three operators in the scalar~\cite{bdm} and in the gauge sector~\cite{Beccaria:2008fi}.

\subsection{Reciprocity-based  Ansatz}

Based on the exceptional number of checks done for a variety of operators and reversing the usual logic, reciprocity can be simply \emph{assumed}, and \emph{used} as a tool to
reduce the number of unknown coefficients in the  standard Ansatz based on the maximum
transcendentality principle to be solved via Bethe equations. 

To see how this procedure can be used in practice let us consider an illustrative example, the two-loop anomalous dimension for twist three scalar operators.
One starts with the following Ansatz of transcendentality $\tau=3$ made of harmonic sums with positive indices and argument $S/2$ (as is the case for twist-three operators made of scalars) 
\begin{equation}
\g_2=a_1 S_3+a_2 \,S_{1,2}+a_3\,S_{2,1}+a_4\,S_{1,1,1}\,.
\end{equation}
The corresponding kernel  has the following form in the canonical basis
\begin{equation}
\label{P_4}
\cP_2=\gamma_4-\frac{1}{4}\g_2\g_2'\equiv(a_1-16) S_3+(a_2+16) S_{1,2}+(a_3+16)
S_{2,1}-16\,\zeta_2\, S_1+a_4S_{1,1,1}\,,
\end{equation}
and when rewritten in terms of the 
$\underline\Omega$ basis the result is
\begin{equation}
\label{P4bis}
\cP_4=c_1\,\underline{\Omega_{1}}+c_3\,\underline{\Omega_{3}}+c_{1,2}\,\underline{\Omega_{1,2}}+c_{2,1}\,\underline{\Omega_{2,1}}+c_{1,1,1}\,\underline{
  \Omega_{1,1,1}}
+{\rm const}\,,
\end{equation}
where the $c_i$ are linear combinations of the coefficients $a_i$. 
The combinations 
$\underline{\Omega_{1}},\,\underline{\Omega_{3}},\,\,\underline{\Omega_{1,1,1}}$ 
are all reciprocity respecting, according to the above theorem.
Imposing reciprocity 
on $P_2$ implies the vanishing of the coefficients of those $\underline\Omega$ 
with wrong parity, namely 
\be
c_{1,2}=a_2+16+\frac{a_4}{2}=0,~~~~~~~~~~~~~~~c_{2,1}=a_3+16+\frac{a_4}{2}=0\,.
\ee
This leads to the conditions $a_3=a_2$ and $a_4=-2(16+a_2)$, that are indeed satisfied 
by the known two-loop expression for the anomalous dimension~\cite{Beccaria:2007cn,Kotikov:2007cy}. Thus, reciprocity has determined $2$ of the $4$ unknown 
coefficients in the initial Ansatz for the anomalous dimension~\footnote{The coefficient $a_4$ 
has only been kept to show the exact number of constraints coming from reciprocity. It could have 
been set to zero from the beginning because at large $M$ the term $S_{1,1,1}\sim \log^3M$ is not 
compatible with the universal leading logarithmic behavior (cusp anomaly).}.
This procedure was used in~\cite{BFLZ} to deduce the five loop
asymptotic part of the anomalous dimension for twist three scalar operators. At this loop order, starting with a linear combination of harmonic sums of
transcendentality $\tau=2n-1=9$ one finds in principle $256$ terms which potentially
contribute to the anomalous dimension. Fitting numerically all the
coefficients, that should come out in exact (rational) form, is rather hard due to computational limitation.
Imposing reciprocity one obtains instead an over-determined set of linear
equations, which is solvable~\footnote{We should stress, however, that reciprocity as an assumptions only acts as a computational tool. As usual in such kind of conjectures, there is a powerful numerical test that can be applied to any guesswork, and the closed formulas presented in~\cite{BFLZ} have been always double checked numerically as solutions of the Bethe equations.}. In the same paper  the leading wrapping
correction has been computed, which turns out to be \emph{separately} reciprocity respecting. We recall that the result based
on this assumption has been later confirmed by a purely field-theoretical calculation~\cite{Fiamberti:2009jw}.

A similar reciprocity-based Ansatz was used in  was also adopted  in~\cite{5loopstwist2} to derive the five-loop calculations for the anomalous dimensions of twist-two operators (see Section 4.1 point 1. above).

\subsection{Summary of weak-coupling reciprocity tests}

The successful application of the methods that we have just illustrated proves that 
the reciprocity property of ${\cal N}=4$ SYM  has a wider range of validity
than expected. It is confirmed at higher loops for the twist-2 universal multiplet and is also 
valid for twist-3 operators built with elementary fields of any conformal spin.
Table~\ref{weakres} summarizes the present status of weak coupling tests. 
 \begin{table}[ht]
\begin{center}
\begin{tabular}{|c|c|c|l|}
\hline
 $\cal O$ & \# loops & wrapping & reciprocity  \\
\hline
$\langle \varphi\varphi\rangle$, $\langle \psi\psi\rangle$, $\langle AA \rangle$ & 5 & yes & $\surd$ \\$\langle \varphi\varphi\varphi\rangle$  & 5 & yes & $\surd$
\\$\langle\psi\psi\psi\rangle$ & 5 & yes & $\surd$ \\
$\langle AAA\rangle$ & 4 & no & $\surd$ (ABA) \\
\hline
\end{tabular}
\caption{Status of weak coupling reciprocity on \emph{minimal} dimensions for twist operators}
\label{weakres}
\end{center}
\end{table}

The results about the universal twist two supermultiplet  (first row in the table) are a consequence of the four loop check (ABA \emph{and} wrapping contributions) in the scalar sector~\cite{Beccaria:2009vt}, of the five loop result of~\cite{5loopstwist2} and of the fact that the constant shift in the spin that relates the anomalous dimensions in the supermultiplet as in (\ref{twist2universal}) doesn't affect their large spin expansion properties, which are at the basis of the reciprocity. With the same motivation and  due to (\ref{gauginoshift}), reciprocity holds with the same features for twist-thee operators built out of gauginos (third row in the table). For the twist-three scalar sector (second row in the table), reciprocity has been proved up to four loops in~\cite{bdm}, and is present separately both in the asymptotic (trivially) and in the wrapping contribution of the  five loop result of~\cite{BFLZ}. Reciprocity for twist-three gauge operators has been proved at three~\cite{Beccaria:2007pb} and at four loops~\cite{Beccaria:2008fi} for the asymptotic part of the anomalous dimension (last row in the table). 

 Let us note that anomalous dimensions of operators with twist higher than two occupy a band~\cite{Belitsky:2006en}, the lower bound of which is the  \emph{minimal} dimension for given $S$ and $J$. 
 Every successful check of reciprocity has been performed at weak coupling only for minimal anomalous dimensions, while in fact anomalous dimensions of operators with twist higher than two with trajectories close to the upper boundary of the band \emph{do not} respect reciprocity, as seen in the twist-three case at weak coupling in~\cite{bkp}.  
However, it is interesting that a relation like (\ref{leadinglogs}) also holds for such excited trajectories \ci{bkp}~\footnote{This is also what we shall see at strong coupling 
on the example of the spiky strings.}.

A brief discussion of further results concerning reciprocity properties of higher conserved charges is contained in App.~(\ref{app:higher}). The extension of the analysis to ABJM models~\cite{Aharony:2008ug} has also been investigated 
and is illustrated in App.~(\ref{app:abjm}).

\section{Reciprocity at strong coupling: semiclassical strings  in $\ads$}

\bigskip

The analysis of the reciprocity property in the strong coupling regime of $\mathcal{N}=4$ SYM is performed by making use of the AdS/CFT correspondence, namely considering  energies of the ``semiclassical string states" which are believed to be dual to the quasipartonic operators~\cite{tseytlinreview}. 
The string states one is referring are solitonic solutions of the string equations of motion carrying a finite 2-d energy that can be expressed in terms of other charges (spins), and the standard semiclassical expansion refers to the energies of strings in $AdS_5\times S^5$ having large quantum numbers and thus dual to ``long'' SYM operators with large canonical dimensions. 

In the following, we will study reciprocity at the level of the energy in the two  cases of folded string and spiky strings, extending the analysis at one loop in the semiclassical expansion for the folded string solution. We will then discuss a generalization of reciprocity at the level of the eigenvalues of the first few commuting charges defined in~\cite{Arutyunov:2003rg}.

 It is of interest to recall  that in such analysis, neither we will  explicitly refer to the classical integrability of the string sigma model~\cite{Bena:2003wd}, nor to the semiclassical approach directly relying on such classical general finite gap description\cite{kaz,gr1}. Interestingly enough,
however, integrability will come up again at the one-loop level 
 via the connection with the integrable, finite-gap, Lam\'e equation~\cite{BDFPT}. 

\subsection{Classical folded string in $AdS_3\times S^1$}

The first and most important example in this sense is the non-trivial rigid string
solution of~\cite{gkp}  describing a folded spinning string  rotating in the $(\rho,\phi)$ plane of $AdS_5$ and moving along the $\varphi$-circle of $S^5$. For this configuration the integrals of motion  are the space-time energy $E=\sl\,\E$ and the two spins  $S=\sl\,{\cal S}$ and $J=\sl\,{\cal J}$ (conserved momenta conjugate to $t$ and to $\phi$, $\varphi$ respectively).  In the full quantum theory $S$ and $J$ should take quantized values. In the semiclassical
approximation we shall consider, however,  their values are assumed to be very large, in such a way that $\cS$ and	${\cal J}$	are finite for $\sl\gg 1$.

The expressions for the ``semiclassical'' energy and spins can be found~\cite{ft1} in terms of the elliptic functions
  $\EE$ and $\KK$ of an
   auxiliary variable $\eta$
\begin{eqnarray}
\label{foldinform1}
 \E&=&\kappa+\frac{\kappa}{\omega}\cS\ ,~~~~~~~~~~~~~
 ~~~~~~~~~~~~~~~~~~~~~~~~~~~~~~~~~~~~~~~
\frac{\omega^2-{\cal J}^2}{\kappa^2-{\cal J}^2}\equiv1+\eta\, ,
\\
\label{foldinform2}
\cS&=&\frac{2\pi\, \omega\,\sqrt{\eta}}{\sqrt{\k^2-\J^2}}\,\left[\EE\left(-
\textstyle{\frac{1}{\eta}}\right)-
\KK\left(-\textstyle{\frac{1}{\eta}}\right)\right],~~~~~~~~~~\sqrt{\kappa^2-{\cal J}^2}=\frac{2}{\pi\,\sqrt{\eta}}\KK\left(-\textstyle{\frac{1}{\eta}}\right)
\end{eqnarray}
Here $\k$ and $\omega$ (or $\eta$)  are parameters of the classical solution which
 should we eliminated to find $\E$ as a function of $\cS$ and $\J$.
  
 To find the energy in terms of the spin one is to solve for  $\eta$.
Here we are interested in the large spin expansion which corresponds
to the long string limit (when the string   ends are close to
 the boundary of $AdS_5$).
For such long string  one has $\eta\to 0$.

In the limit in which the $S^5$ momentum $J$ of the string state can be ignored,  solving for $\cS$ in (\ref{foldinform2}) for small  $\eta$  and substituting it
  into the first of (\ref{foldinform2}), one finds for $\E$  as a function of ${\cal S}$ the expansion
\begin{eqnarray}
&&{\cal E}= \cS+\frac{\log\bar \cS-1}{\pi} + \frac{\log\bar \cS-1}{2\,\pi^2\,\cS}
-\frac{2\log^2\bar \cS-9\log\bar \cS+5}{16\,\pi^3\cS^2} \no\\
&& \ \ \ \ \ \ \ \ \ \ \
+\ \frac{2\log^3\bar\cS-18\log^2\bar\cS+33\log\bar\cS-14}{48\,\pi^4\,\cS^3}+...\ , \ \ \ \ \ \  \ \ \
\bar \cS\equiv 8\,\pi\,\cS \ ,
\label{EmS1}
\end{eqnarray}

In the case in which the $S^5$ angular momentum of the string is not negligible compared to $S$, 
i.e. when the string state  is dual to an operator
 with large spin $S$    and large twist $J$, one can work out analogous expansions.
We will be  interested in large $\cS$  expansion   with   $\cS \gg \J$
since only  in this case the expansions  like (\ref{genexp}), i.e. going
  in the  inverse powers of $\cS$ with the
coefficients being polynomials  in $\log \cS$,  will apply (see also \cite{ft1,bk}).
 
In the large $\cS\gg \J$ or  \emph{long string} limit, when  $\eta\ll 1$,
 one should distinguish between  ``small'' or  ``large''  $\J$ cases \cite{ft1,ftt}.
In the ``slow long string" approximation  (corresponding to taking $\cS $ to be large  with
$\ell \equiv { \frac{\J }{ \log \cS}} $ fixed and then expanding in powers of $\ell$)
the leading  terms in  the semiclassical energy read (cf. \ref{EmS1})
 \begin{eqnarray}\label{enslow}
\E-\cS-\J&\approx&\frac{1}{\pi} (\log{\tS}-1)
+\frac{\pi\,\J^2}{2\,\log\tS}
-\frac{\pi^3\,\J^4}{8\,\log^3\tS}\big(1-\frac{1}{\,\log\tS}\big)+...\\\nonumber
&&~~~~+\frac{4 }{\tS}\Big[{ 1 \ov \pi} (\log{\tS}-1)
+\frac{\pi\,\J^2}{2\log^2{\tS}}
-\frac{3\pi^3\,\J^4}{4\log^4{\tS}}\big(1-\frac{2}{3\,\log\tS}\big)+...\Big]\\\nonumber
&&~~~~-\frac{4}{\tS^2}\Big[{1\ov \pi} (2\log^2{\tS}-9\log{\tS}+5)+\pi\,\J^2
\Big(1+\frac{3}{2\log{\tS}}
-\frac{1}{\log^2{\tS}}-\frac{2}{\log^3{\tS}}\Big)+...\Big]
\end{eqnarray}
where  $\tS\equiv 8\pi\cS$,  and dots stand  for higher order
  corrections depending on  $\J$. 
In the case of ``fast long string'', when
 $\log\cS\ll\J\ll \cS$, the corrections to the energy read
\begin{eqnarray}\nonumber
&&\!\!\!\!\!\!\!\!\!
\E-\cS-\J\approx\frac{1}{\pi^2\,\J}\Big[{ 1\ov 2} {\log^2\hS}-\log\hS+\frac{4\log\hS}{\hS}+
\frac{4}{\hS^2}\big(-2\log\hS+1+\frac{3}{\log\hS}+\frac{2}{\log^2\hS}+...\big)+...\Big]\\\label{enfast}
&&
~~~~~~~~~~+\frac{1}{\pi^4\J^3}\Big[-\frac{\log^4\hS}{8}-\frac{2}{\hS}\big(3\log^2\hS+
\log\hS+1+\frac{1}{\log\hS}+\frac{1}{\log^2\hS}+...\big)\\\nonumber
&&~~~~~~~~~~~~~~~~~~~~~~~ -\frac{2}{\hS^2}\big(2\log^3\hS-19\log^2\hS+11\log\hS+
13+\frac{13}{\log\hS}+\frac{11}{\log^2\hS}+...\big)+...\Big]
\end{eqnarray}
where $\hS\equiv
\frac{8S}{J} =\frac{8\cS}{\J}\gg 1$. Dots in the square brackets
 indicate corrections in $1/\hS$, corrections in  $1/\log\hS$ can
  be added in the round brackets and  terms like $ \log(\log\hS)$
  have been neglected.

With the large spin expansions (\ref{EmS1})-(\ref{enfast}) at hand, we first observe a general  agreement in  the {\it structure} of the large $S$ expansion as found
in perturbative string theory  and in perturbative gauge theory, see  (\ref{genexp}).  This agreement is non-trivial  since  the gauge-theory and string-theory
perturbative expansions are organized differently:
the gauge-theory limit is to expand in small $\l$ at fixed $S$
 and  then expand the $\l^n$ coefficients in large $S$,
while the semiclassical string-theory limit is to expand in large $\l$  with fixed $\cS =
{ S \ov \sl}$  and then expand the $ 1 \ov (\sl)^n$  terms in $E$  in large
$\cS$.
Even assuming these limits commute (which so far appears to be verified only
 for  the leading universal  $\log S$ term) the reason for the  validity of the
 functional relation (\ref{Nonlinear})  and, moreover, of  the reciprocity property  (\ref{Parity}) is 
  obscure on the semiclassical string theory side.

We can furthermore study the compatibility of the expansions found with the functional relation (\ref{Nonlinear}). In particular, 
the coefficients of the leading $ ({\log \cS \ov \cS})^m $
 terms in \rf{EmS1} happen, indeed, to be  consistent
  with the equation (\ref{leadinglogs}),
    with the leading term in the function
    $\rm f$ being simply the logarithm
\be\label{lead}
E-S=
 \frac{\sqrt{\l}}{\pi}\log\Big[S+\frac{1}{2}\frac{\sqrt{\l}}{\,\pi}\log  S +...\Big]+... \ .
\ee
The same it's true for the expression (\ref{enslow}), where the  leading terms  in the expression of  (\ref{EmS1}) dominate in the limit when $\frac{\J^2}{\log\cS} \ll { \log \cS \ov \cS }$.
In the case of the expansion  (\ref{enfast}), the leading terms can be summed up as~ \cite{Belitsky:2006en}
\be
\E- \cS = \sqrt{ \J ^2 + {1 \ov \pi^2} \log^2 { 8 \cS \ov \J }}  + ...\ ,  \ee
where ${  \log \cS \ov \J } \ll 1 $ plays the role of  an expansion parameter.
Notice that in contrast to the  slow long string case where  the expansion
(\ref{enslow}) has the same structure  as in (\ref{genexp}), in the fast long string
case (\ref{enfast}) one gets  higher powers of $\log \cS$  not suppressed by $\cS$ ~\footnote{For this kind of discrepancy with the weak-coupling behavior one would in general need a resummation of the type discussed at the end of this Section.}. Neverthless, the reciprocity property can be successfully checked as we explain below.

It is then possible to proceed as follows with the analysis of reciprocity. If one  identifies  the energy $E$,  the angular momenta $S$ and $J$ of a string rotating in a plane  in global $AdS_5$
with dimension, Lorentz spin and twist  of the  gauge theory quasipartonic
operators,   the functional relation
(\ref{Nonlinear})  would then   imply   that  the anomalous dimension  
 should be  a function (that we rename as $\rm f$ in this strong coupling context) of  itself as in
   \be  
\g=
E-S-J = {\rm f} \,\Big( S+\frac{1}{2} \g\Big)   \ . 
\label{faw} 
\ee
To take into account the peculiarity of the  string
 semiclassical perturbation theory,  where   all non-zero charges are
automatically large at large $\l$, we shall use the semiclassical analogs  $\tg=\frac{\g}{\sl}$, $\tilde {\rm f}=\frac{\rm f}{\sl}$ of the function appearing in (\ref{faw}), checking therefore whether the function $\tf$ defined in
\be\label{Nonlineartilde}
\tg=\tf\,\Big(\cS+\frac{1}{2}\,\tg \Big)~~~~~~~~~{\rm as}~~~~~~~~~\tf=\sum_{k=1}^\infty\frac{1}{k!}\,\Big(-\frac{1}{2}\frac{d}{d\cS}\Big)^{k-1}[\tg]^k
\ee
admits an expansion in even negative powers of the semiclassical analog ${\cal C}=\frac{C}{\sl}$ of the Casimir in (\ref{Casimir}). This will be ${\cal  C }\equiv\cS$ in the case of a folded string rotating only in $AdS$, and ${\cal C}\equiv \cS+\frac{1}{2}\,\J$ in the case of the folded string rotating in $AdS_5$ with non-zero angular momentum in $S^5$~\footnote{The choice for this case of $\ell=\frac{1}{2}$ in the semiclassical version of (\ref{Casimir}) follows from the fact that the non-zero $R$-charge for classical bosonic solutions automatically selects the $sl(2)$ sector identified in fact by $\ell=\frac{1}{2}$. }.

Specifically, for the $AdS$ folded string,  the large $\cS$ expansion of the function ${\tilde {\rm f}}$ (its leading term in the strong-coupling limit)
 is much simpler than that of   the anomalous dimension $E-S$ in  (\ref{EmS1})
and contains only {\it even} powers of  $C^{-1} \sim \cS^{-1}$ 
\be\label{P01} 
{\tilde {\rm f}}(\cS)={1 \ov \pi} \Big[ \log \bar \cS-1
+\frac{\log \bar \cS+1}{16 \pi ^2\cS^2 }  + {\cal O}\left({1\ov \cS^4}\right)\Big] +  {\cal O}({1 \ov \sl})  \ .
\ee

\bigskip

A more systematic analysis  of the reciprocity  (parity invariance)
 property  of the function $\tf$ is possible with the  help of   an integral
 representation for it.
Using that (\ref{Nonlineartilde}) implies
$\tf(\cS')= \tg\left(\cS'-\textstyle{\frac{1}{2}}\tf(\cS')\right)$, where $\cS'= \cS + \frac{1}{2} \tg (\cS)$,   $\tg(\cS) = \E - \cS$,
and renaming $\cS' \to \cS$
we have
\be\label{LB}
\tf(\cS)=\frac{1}{2\pi \,i}\oint_{\Gamma} d\omega \ \tg(\omega)\ \frac{1+\frac{1}{2} {\tg'(\omega)}}{
\omega-\cS+\frac{1}{2} {\tg(\omega)}}\ ,
\ee
where the contour $\Gamma$ encircles
 the pole of the integrand and prime stands for derivative.\footnote{The expression that
  multiplies $\tg$ in the integrand has residue $1$, so that
   the integral is $\tg$ evaluated at the
   pole $\omega=\cS-\frac{1}{2} {\tg}$.
   Then defining $x=\cS-\frac{1}{2}\tf(\cS)$
    we have   $2 \cS-2x=\tg$ which coincides with the
     equation for the pole with $x=\omega$.}
It is  natural to replace  the variable $\omega$ in (\ref{LB}) with
the expression (\ref{foldinform2}) for the semiclassical spin $\cS(\eta)$
\be\label{B2}
\tf(\cS)=\frac{1}{2\pi \,i}\oint_\Gamma\,d\eta\, \tg(\eta)\,\frac{{\tilde s}'(\eta)}{{\tilde s}(\eta)-\cS} \ ,
\ee
where ${\tilde s}(\eta)\equiv\cS(\eta)+\frac{1}{2}\tg(\eta)= \frac{1}{2} (\E + \cS)$ is
the renormalized ``conformal spin'', see formula (\ref{ko}), expressed  in terms of the
 semiclassical quantities. The  integral then gives
 the function $\tg$ evaluated at  the zero of the denominator;
 this  is the same as the  statement that the anomalous dimension as a function of
  the Lorentz spin is, effectively, a function of
  the conformal spin ${\tilde s}$.

To verify the reciprocity property  of the function $\tf(\cS)$
 in  (\ref{B2}) it is useful to redefine  the variable $\eta$
 as\footnote{This choice is not unique.
 An analogous  transformation was used in~\cite{bk}.}
   $\eta\rightarrow -1+16\eta+\sqrt{1+256\,\eta^2}$
   and examine the large $\cS$ or small $\eta$ limit of the expressions.
One finds  that  $\tg(\eta)$  is a series in even powers of $\eta$
\be\label{ven}
\tg(\eta)=-\frac{1+\log\eta}{\pi}+\frac{4(\log\eta+12)}{\pi}\eta^2-\frac{6(62\log\eta+777)}{\pi}\eta^4+... \ ,
\ee
while the expression for the conformal spin runs in odd powers of $\eta$
\be\label{nodd}
{\tilde s}(\eta)=\frac{1}{8\pi\eta}+\frac{11+2\log\eta}{2\pi}\eta -\frac{877+92\log\eta}{2\pi}\eta^3 +... \ .
\ee
From the equation for the pole of the integrand in (\ref{B2}),
 ${\tilde s}-\cS=0$, one can find  the parameter $\eta$ in terms of
 the spin $\cS$,  concluding that it is given by a power series in  odd negative powers of $ \cS$.
As a result,  $\tf(\cS)$, which is  same as $\tg(\eta)$ evaluated at  the pole, should
 also run only in even negative powers   $\C \equiv \cS$.

\bigskip

Coming to the case of the folded $AdS_5$ string with non-zero angular momentum in $S^5$, one may  again make  use of the integral representation for the
 functional relation as in (\ref{LB}).
The discussion will  apply to  both the  ``slow'' and the ``fast'' long string limits.
Here the renormalized ``conformal spin''
is  ${\tilde s}=\frac{1}{2} ( \cS + \E)=\cS+\frac{1}{2}\J+\frac{1}{2} \tg $, and we anticipated that
 the semiclassical value of the
 Casimir operator  is
 $\C\equiv \cS + \frac{1}{2} \J$.
Then
 the integral in (\ref{B2}) can be written as
\be\label{LB2}
\tf(\C)=\frac{1}{2\pi \,i}\oint_\Gamma\,d\eta\, \ \tg(\eta)\,\frac{{\tilde s}'(\eta)}
{{\tilde s}(\eta)-\C}\ , \ \ \ \ \ \ \ \ {\tilde s}(\eta)=\cS(\eta)+\frac{1}{2} \tg(\eta)   \ .
\ee
After a redefinition  of $\eta$ one can then show   that
 the  expansion of $\tf$ in large $\C$
  runs only in even negative powers of $\C$ (see Appendix D of~\cite{BFTT}).
In the  kinematic region of ``fast'' long strings,  with
$1 \ll \log\cS\ll\J\ll\cS$,  this parity invariance  property
was already demonstrated in a closely related way
 in~\cite{bk}.

Notice that to establish a relation to the definition of reciprocity
 in weakly coupled gauge theory expansion with finite twist 
 one would need  to consider the case
 of semiclassical $(S,J)$ string and then resum the series  for its energy
 (both in $J$ and in $\sqrt{\l}$)
 so that the limit  of finite $J$  would  make sense. This is due  to the subtlety of semiclassical string expansion, again because  all non-zero charges are
automatically large at large $\l$ and, for example,  the case of finite twist $J=2,3,...$
can not be distinguished from the formal case of $J=0$.
It is usually assumed
that the folded string in $AdS_5$ with zero angular momentum in $S^5$
describes an operator  of small twist, but that can be
 $J=2$ or $J=3$, etc.

\subsection{Spiky strings in $AdS_5$ and classical violation of reciprocity}

It is interesting to mention a relevant example in which reciprocity is 
 \emph{violated} already at classical level. This is the case of the spiky spinning string  in $AdS_5$~\cite{kru}, the integrals of motion  are the energy, the spin (angular momentum in $AdS_5$) and  the  difference between the position of the spike and of the middle of the valley between the two spikes, $\Delta \theta=\frac{\pi}{n}$, expressed in terms of the number of the spikes $n$. Also in this case it is possible to perform a large spin expansion, corresponding to the ends of the spikes approaching the boundary of $AdS_5$, which reads~\cite{BFTT}
\begin{eqnarray}\label{EmSspikes}
&&\E-\cS=\frac{n}{2\pi}\Big[ \log\tS  + p_1   
+\frac{4}{\tS}\left(\log\tS + p_2 \right) -\frac{4}{\tS^2}\left(2\log^2\tS  + p_3 \log\tS+ p_4  \right) \no \\
&&\ \ \ \ \ \ \ \ \ \ \ \ \ \ \ \ \ \ \ \
+\ \frac{32}{3\,\tS^3}\left(2\log^3\tS  + p_5  \log^2\tS+ p_6  \log\tS  + p_7   \right) +...\Big]\ ,
\end{eqnarray}
where   $\tS={16\,\pi\ov n}\cS $ and
\begin{eqnarray}\label{opr}
&& p_1  = -1+\log\sin\frac{\pi}{n} \ ,~~~~~~~~~~~~~~~~~
p_2= -1+\log\sin\frac{\pi}{n}+\frac{\pi(n-2)}{2n}\cot\frac{\pi}{n}\ ,\\
&&p_3= - 10 +  \frac{2\pi (n-2)}{n} \cot \frac{\pi }{n}  - 2  \cot^2\frac{\pi }{n}  
 -4  \log \csc \frac{\pi }{n}   +  \csc^2\frac{\pi}{n} ,
 \\
&& p_4=  6 - \csc ^2\frac{\pi}{n}+ \frac{\pi ^2 (n-2)^2}{2 n^2} -\frac{4\pi (n-2)}{n} \cot \frac{\pi }{n}  + 
\cot^2\frac{\pi }{n} \Big[\frac{\pi ^2 (n-2)^2}{n^2} +1\Big] \no \\
&&\ \ \ \ \ \ \ \ + \
\log \csc \frac{\pi }{n}\ \Big[2 \cot^2\frac{\pi }{n}- \frac{2 \pi(n-2)}{n}\cot\frac{\pi }{n}
- \csc ^2\frac{\pi }{n} +2 \log \csc \frac{\pi}{n}+10\Big]
\ ,\\
&& p_5= -18 + {\cal O}(n-2) \ , \ \ \ \ \
p_6= 33 + {\cal O}(n-2) \ , \ \ \ \ \
p_7=  -14 + {\cal O}(n-2) \ . 
\end{eqnarray}
It is easy to check  that (\ref{EmSspikes}) coincides with the energy (\ref{EmS1}) for the folded string in $AdS_5$   when $n=2$.
Retaining in (\ref{EmSspikes}) only the dominant contributions at  each order of
 the above expansion we obtain
\bea
\E-\cS= \frac{\,n}{2\pi}\log\cS  +\frac{n^2}{8\,\pi^2\cS} \log\cS
-\frac{n^3}{64\,\pi^3\cS^2} \log^2{\cS}    +\frac{n^4}
{384\,\pi^4\cS^3}  \log^3\cS +...\ .
\eea
This  may be rewritten as
\be\label{spk}
E-S= \frac{\sqrt{\l}\,n}{2\pi}\log\Big[S +\frac{1}{2}\frac{\sqrt{\l}\,n}{2\,\pi}
\log S \Big]+... \ ,
\ee
implying that the functional relation is satisfied (cf. \ref{leadinglogs}).

Evaluating now the analog of the function ${\tf}(\cS)$ in (\ref{P01}), one finds the following expansion
\be\label{Pspi}
\tf(\cS)=\frac{n}{2\pi}\Big[\log\tS  + q_1 
+ \frac{q_2}{\tS} +\frac{1}{\tS^2}(q_3\,\log\tS+q_4)  +\frac{1}{\tS^3}(q_5\,\log\tS+q_6) ...\Big]+... \ ,
\ee
where 
\begin{eqnarray}
  q_1&=&-1+\log\sin\frac{\pi}{n}\ , \ \ \ \ \ \ \ q_2= \frac{2 \pi (n-2)}{n} \cot\frac{\pi}{n}\ ,\ \
  \   \ \ \ 
  q_3=4\csc^2\frac{\pi}{n}\ ,\\  
  q_4&=&4+2 \pi^2\,\big(\frac{n-2}{n}\big)^2\,\big(1-2\csc^2\frac {\pi}{n}\big)+ 4\log\sin\frac{\pi}{n}\csc^2\frac{\pi}{n}\ ,\\
   q_5&=& {\cal O}(n-2)\ , \ \ \ \ \ \ \ \ \ \ \ \ q_6= {\cal O}(n-2)\ , 
   \label{suu}
\end{eqnarray}
with $q_5,q_6$ are non-zero for $n \not=2$.
The expansion (\ref{Pspi}),  even if considerably simpler 
compared  to the energy (\ref{EmSspikes}), is  not parity invariant under $\cS\to-\cS$.
 The parity invariance is restored  in the case of the folded
   string when $n=2$, where indeed (\ref{Pspi}) coincides with (\ref{P01}).

This breakdown of parity invariance for a string with  $n > 2$ spikes
is actually not only non surprising, but expected. In fact, such spiky string  should  correspond  to
an operator with \emph{non-minimal} anomalous dimension for a given spin,
 while the reciprocity  was checked at weak coupling only for the minimal
 anomalous dimensions.
Indeed and as already mentioned, anomalous dimensions of operators
 of twist higher than two with trajectories close to the
 upper boundary of the band present features
 completely analog to the one seen here,
 in that they satisfy (\ref{leadinglogs}) while violating reciprocity~\cite{bkp}~\footnote{It is interesting that our strong-coupling result (\ref{spk}), (\ref{Pspi}) has close similarity with  weak-coupling one found for $n=3$  in \ci{bkp}: the functional 
relation (\ref{leadinglogs}) is still satisfied,  and the parity invariance is broken at level $1/S$.}.

\subsection{Reciprocity in string perturbation theory}

The observation that  reciprocity  holds at 1-loop in string semiclassical expansion, first made  in~\cite{BFTT}, has been confirmed and extended in~\cite{BDFPT} in the case of a folded string rotating in $AdS$. 
The standard string semiclassical approximation is based on expanding the energy $E$  in large $\sl$ with ${\cal S} =S/\sl$ kept fixed,
\be\label{semicl}
E = E\Big(\frac{S}{\sl},\sl\Big) = \sl\,\E_0(\cS ) + \E_1(\cS ) +\frac{1}{\sl}\,\E_2(\cS ) + ... 
 \ee
where $\E_0$, the classical energy, coincides with (\ref{foldinform1}) , and $\E_1$, $\E_2$  are the 1-loop and 2-loop energies.  translates into an analog semiclassical expansion within the relation (\ref{Nonlineartilde}). Namely, the ``anomalous dimension'' can be written
\be
\tg=\tg_0+\frac{1}{\sl}\,\tg_1+...,~~~~~{\rm where}~~~~~\tg_0=\E_0(\cS)-\cS,~~~~~~~~\tg_1=\E_1(\cS)
\ee
from which the function $\tf$ defined by (\ref{Nonlineartilde}) can be determined as in
\be
{\tf} = {\tf}_0 + \frac{1}{\sqrt\lambda}\,{\tf}_1 + \cdots,
\ee
with
\ba
{\tf}_0 = \sum_{k=1}^\infty\frac{1}{k!}\left(-\frac{1}{2}\frac{d}{d{\cal S}}
\right)^{k-1}[\tg_0]^k, ~~~~~~~~~~~~~~~
{\tf}_1  = \sum_{k=1}^\infty\frac{1}{k!}\left(-\frac{1}{2}\frac{d}{d{\cal S}}
\right)^{k-1}[k\,\tg_0^{k-1}\,\tg_1].
\ea
 It is a recent achievement~\cite{BDFPT}, due to the observation that the semiclassical fluctuation problem is governed by standard single-gap Lam\'{e} operators,  the possibility to write down an analytic exact expression for the relevant functional determinants.
From the exact one-loop energy $\E_1\equiv\g_1$ that can be written in terms of them, it has been possible to extract the following expression for its large spin (small $\eta$) 
expansion~\footnote{See also the comment at the end of this section.
}
\ba\no
\tg_1 &=& \frac{\kappa_0}{\kappa} \,\Big[\big(c_{01}\,\kappa_0 + c_{00} +
\frac{c_{_{0,-1}}}{\kappa_0}\big)+ \,\big(c_{11}\,\k_0+c_{10} + 
\frac{c_{_{1,-1}}}{\kappa_0}\big) \eta
+  \\\label{E1large}
& & \ \ \ + \ \big(c_{21}\,\kappa_0 + c_{20} + 
\frac{c_{_{2,-1}}}{\kappa_0}\big)\eta^2 + 
 \,
\big(c_{31}\,\kappa_0 + c_{30} + \frac{c_{_{3,-1}}}{\kappa_0}\big)\eta^3 +{\cal O}(\eta^4) \Big] \ ,
\label{ordereta3}
\ea
where $\k_0=\frac{1}{\pi}\log\frac{16}{\eta}$ and the explicit values for the coefficients are 
\ba\label{coefficientsfirst}
c_{01} &=& -3   \log 2\ , ~~~~~~~~~~~~
c_{00} = 1 +{6\ov \pi} \log 2 \, ~~~~~~~~~~~~
c_{_{0,-1}} = -\frac{5}{12}\ , \\
c_{11} &=& 0 \ ,~~~~~~~~~~~~~~~~~~~~~
c_{10} = -{3\ov \pi} \log 2 \,~~~~~~~~~~~~~~~
c_{_{1,-1}}=\frac{1}{2\pi }+\frac{3\,\log2}{\pi^2}\ ,  \\
c_{21} &=& -\frac{\pi }{32}-\frac{3}{32}   \log 2\ ,~~~~~~~
c_{20} = \frac{1 }{16}+\frac{39 \log 2}{32\pi}\ ,~~~~~~~
c_{_{2,-1}} = -\frac{13}{64\pi}-\frac{63 \log 2}{32\pi^2 }\ ,\\\label{coefficientslast}
c_{31} &=& \frac{\pi }{32}+\frac{3 }{32}  \log 2\ ,~~~~~~~~~~
c_{30} = -\frac{3  }{32}-\frac{13 \log 2}{16\pi }\ ,~~~~~~
c_{_{3,-1}} = \frac{29}{192\pi}+\frac{85 \log 2}{64 \pi^2 }\ .
\ea
Solving for the parameter $\eta$ explicitly in terms of $\cS$, the first few terms in (\ref{ordereta3}) read
\begin{eqnarray}\no
&&\!\!\!\!\!\!\!\!\!\!\!\!\!\!
\tg_1= -\frac{3 \log2}{\pi } \log\bar{\cS}+\frac{\pi+6 \log2}{\pi
   }-\frac{5 \pi }{12 \log \bar{\cS}}+\\
&&   -\frac{1}{\bar\cS}\Big[\frac{24 \log2}{\pi } \log \bar{\cS}-\frac{4\pi+36 \log2}{\pi
   }+\frac{5 \pi }{3 \log^2\bar\cS }\Big]+{\cal O}\left(\frac{1}{\bar{\cS}^2}\right)\\\label{E1largespin}
\end{eqnarray}
with $ \bar{\cS}=8\,\pi\,\cS$.
Working out ${\tf}_1$ and looking at all terms which are odd under ${\cal S}\to 
-{\cal S}$ we find that they vanish  if the following reciprocity constraints hold
\ba
c_{10} &=& \frac{1}{\pi}\,c_{01}\ , \ \ \ \ \ \ \ \
c_{_{1,-1}} = \frac{1}{2\pi}\,c_{00}\ , \ \ \ \ \ \ \ 
c_{31} = -c_{21}\ , \\
c_{30} &=&-c_{20}-\frac{1}{6\pi} c_{01}+\frac{1}{\pi}\,c_{21}\ ,  \\
c_{_{3,-1}} &=& -c_{_{2,-1}}+\frac{1}{4\pi^2}\,c_{01}-\frac{1}{12\pi} c_{00}+\frac{1}{2\pi}\,c_{20}.
\ea
With the list of  explicit coefficients  above (\ref{coefficientsfirst})-(\ref{coefficientslast}),
 these relations are indeed satisfied~\cite{BDFPT}. 
 
As we remarked, the expression of the one-loop energy derived in~\cite{BDFPT} is exact. However, its expansion 
at large spin is quite non trivial. It contains a part which can be computed 
analytically in closed form  and a  reminder, starting at order ${\cal O}(\eta^{2})$, which is known  (as yet) only in implicit form.  It is the large spin expansion of the first contribution, namely formula (\ref{E1large}) above, which turns out to be separately reciprocity respecting~\footnote{This situation is for certain aspects similar to the \emph{ABA +  wrapping} splitting discussed at weak coupling.}.

 \section{Open problems and perspectives}

From the point of view of AdS/CFT, it is quite important to look for common structures shared by the two sides of the
correspondence. Integrability is certainly one of them. The reciprocity property discussed in this 
Review is another  example. Hence, we believe that it is important to 
pursue its investigation and for this reason we list in this final section some related open problems.

First of all, as remarked in the Introduction, there is no rigorous proof of 
reciprocity neither at weak nor at strong coupling. It would be nice to establish the validity of this (discrete) 
hidden symmetry
by a solid physical argument or, possibly, as a mathematical feature of the integrable structures of AdS/CFT, {\em i.e.} Bethe Ansatz equations, Baxter formalism, or exact $S$-matrix.

Another important issue is the connection between reciprocity and wrapping corrections. The latter
are under intense study and are expected to clarify several  interesting facets of a very non trivial pair of 
integrable models. From this point of view, the observation that reciprocity is separately satisfied by the asymptotic 
Bethe Ansatz predictions as well as from the wrapping corrections is an unsolved puzzle.
As a related problem, reciprocity deserves of course further study  in larger (with rank greater than one) sectors of the theory.

Our final comment concerns the strong coupling regime of the gauge theory, which is string perturbation theory. There are currently two apparently 
alternative formalisms to work out quantum corrections for string configurations in $\ads$. The first is 
standard field-theoretical analysis of the string world-sheet $\sigma$-model. This approach, certainly boosted by integrability,  is a priori independent on it. 
The second method is based on of the algebraic spectral curve which, instead, imposes and exploits integrability from  scratch. Currently, it is not totally clear how to relate the two approaches. The signals of reciprocity that we 
have illustrated in the world-sheet calculations are, in our opinion, a very interesting  check and a challenge for the  spectral curve method.

\section*{Acknowledgments}

We  would like to thank A. V. Belitsky,  Yu. L. Dokshitzer, G. V. Dunne, A.V. Kotikov, T.  Lukowski, G. Marchesini, M. Pawellek, A. Tirziu, A. A. Tseytlin and S. Zieme for the nice collaboration on some of the papers on which this Review is based. V. F. is supported by the Alexander von Humboldt foundation.

\appendix

\section{Harmonic sums}
\label{app:harmonicsums}

The nested harmonic sums  $S_{a_1, \dots, a_\ell}$ are
defined recursively 
\begin{equation}
S_a(S) = \sum_{n=1}^S\frac{\varepsilon_a^n}{n^{|a|}}, ~~~~~~~~~~~
S_{a, \mathbf{b}}(S) = \sum_{n=1}^S\frac{\varepsilon_a^n}{n^{|a|}}\,
S_{\mathbf b}(n),
\end{equation}
where $\varepsilon_a = +1(-1)$ if $a\ge 0$ ($a<0$).
The \emph{depth}  of a given sum $S_\mathbf{a} = S_{a_1, \dots, a_\ell}$ is
defined by the integer $\ell$, while its \emph{transcendentality}  is the sum
$|\mathbf{a}|= |a_1| + \cdots + |a_n|$.
The product between harmonic sums can be reduced to linear combinations of
single sums iteratively using the so called shuffle algebra~\cite{Blumlein}; 
\begin{eqnarray}
\label{eq:shuffle}
&& S_{a_1, \dots, a_\ell}(S)\,S_{b_1, \dots, b_k}(S) =\sum_{p=1}^S
\frac{\varepsilon_{a_1}^p}{p^{|a_1|}}\,S_{a_2, \dots, a_\ell}(p)\,S_{b_1,
  \dots, b_k}(p) + \\
&& + \sum_{p=1}^S \frac{\varepsilon_{b_1}^p}{p^{|b_1|}}\,S_{a_1, \dots,
  a_\ell}(p)\,S_{b_2, \dots, b_k}(p) - \sum_{p=1}^S
\frac{\varepsilon_{a_1}^p\,\varepsilon_{b_1}^p}{p^{|a_1|+|b_1|}}\,S_{a_2,
  \dots, a_\ell}(p)\,S_{b_2, \dots, b_k}(p). \nonumber
\end{eqnarray}

{\em   Complementary and subtracted sums}\\

Let $\mathbf{a} = (a_1, \dots, a_\ell)$ be a multi-index. For $a_1\neq 1$, it
is convenient to adopt the concise notation
\begin{equation}
S_\mathbf{a}(\infty)\equiv S^*_\mathbf{a}.
\end{equation}
We define the complementary harmonic sums recursively by $\underline{S_a} = S_a$
and 
\begin{equation}
\underline{S_\mathbf{a}} = S_\mathbf{a}-\sum_{k=1}^{\ell-1} S_{a_1,\dots,
  a_k}\,\underline{S_{a_{k+1},\dots, a_\ell}^*}.
\end{equation}
Note that the definition is ill when $\mathbf{a}$ has some rightmost 1
indices; In this
case, we will treat $S_1^*$ as a formal object in the 
above definition and will set it to zero in the end. 
Since $\underline{S_\mathbf{a}}^* <\infty$ in all remaining cases, it is meaningful
to  introduce the subtracted complementary sums, defined as follows: 
\begin{eqnarray}
\underline{\widehat{S}_\mathbf{a}} &=& \underline{S_\mathbf{a}}
-\underline{S_\mathbf{a}^*}.
\end{eqnarray}
The explicit form of the above definition is
\begin{equation}
\label{subtracted}
{\underline{\widehat
    S_{\mathbf{a}}}}(S)=(-1)^\ell\,\sum_{n_1=S+1}^\infty\frac{\vare_{a_1}^{n_1}}{n_1^{|a_1|}}\,
\sum_{n_2=n_1+1}^\infty\frac{\vare_{a_2}^{n_2}}{n_2^{|a_2|}}\,
\dots\,\sum_{n_\ell=n_{\ell-1}+1}^\infty\frac{\vare_{a_\ell}^{n_\ell}}{n_\ell^{|a_\ell|}}.
\end{equation}

\section{Reciprocity of higher conserved charges}
\label{app:higher}

To the notion of integrability for the spin chains corresponding to ${\cal N}=4$ SYM composite operators
is associated the existence of an infinite tower of commuting charges, in standard notation $\{q_{r}\}_{r\ge 2}$.
The first of them $q_{2}$ is identified with the Hamiltonian of the chain and one refers to 
a hierarchy of {\em conserved} charges. Actually, in our context all the $q_{r}$ are on the same footing and is then natural to extend the analysis of the reciprocity properties to the
full set of conserved charges. An attempt in this direction is the paper~\cite{Beccaria:2009yt} 
where the reader can find more details. Here, we just summarize the main outcomes of that analysis.

In~\cite{Beccaria:2009yt}, a few higher charges in the $\mathfrak{sl}(2)$ subsector are studied.
In the weak coupling regime, the first two non trivial charges $q_{4,6}$ 
have been computed at three and two loops respectively. 

The result of the analysis is that reciprocity is indeed at work.
The definition of the kernel $\cP_r$ (see Eq.~(\ref{Nonlinear})) can be consistently generalized to the 
full tower of charges according to 
\begin{equation}
q_r(S) = \cP_r\,\Big(S + \frac{1}{2} q_2(S)\Big).
\end{equation}
Notice that this definition involves the  renormalized conformal spin $S + \frac{1}{2}\,q_2(S)$
as argument of the kernel, in agreement with light-cone quantization. The naive argument 
$S + \frac{1}{2}\,q_r(S)$ implicitly defines a non reciprocity-respecting kernel.

The strong coupling regime can be explored 
at the classical level considering the first higher charges of the sigma model,
which can be derived from those of 
the $\mathfrak{su}(2)$ sector~\cite{Arutyunov:2003rg}  by analytic
continuation and then analyzed following the same strategy adopted for the energy case.
At this leading order, the parity invariance is satisfied by all the examined charges.

As a final comment, we remark that the  wrapping corrections for the higher charges have 
not been computed yet, even at the leading order. It would be very nice to include them
in the TBA treatment.

\section{Reciprocity and ABJM theory}
\label{app:abjm}

In this Review, we considered $N=4$ SYM duality with string propagation on $\ads$. Actually, 
integrability appears in other instances of the AdS/CFT correspondence. 
In particular the correspondence between the so
called ABJM theory~\cite{Aharony:2008ug} and IIA string on $AdS_4\times \mathbb{CP}^3$ has been recently
widely studied.

 Again, the string model is classically integrable~\cite{Arutyunov:2008if,Stefanski:2008ik,Gomis:2008jt}.
The dual gauge theory is  a ${\cal N}=6$ superconformal theory 
in three dimensions, with $U(N)\times U(N)$ gauge 
group and Chern-Simon action with opposite levels $+k$, $-k$, emerging in the low
energy limit of a theory of $N$ branes  at a $\mathbb{C}^4/\mathbb{Z}_k$ singularity.\\ 

In~\cite{Minahan:2008hf,Bak:2008cp} it has been shown that the dilatation operator for
single trace operators built with 
the scalars of the theory leads to an $SU(4)$ integrable spin
chain, and soon the set of all-loop
Bethe-Ansatz equations for the full $\mathfrak{osp}(2,2|6)$ theory has 
been proposed. Despite that the $\mathcal{N}=4$
SYM and the ABJM theory present a very different structure, one can identify a 
$\mathfrak{sl}(2)$~\cite{Gromov:2008qe,Zwiebel:2009vb} sector in the ABJM theory and
the relative all-loop conjectured Bethe equations show strong similarities
with the SYM case.
Thus, it is a interesting task try to investigate to which extent one can recover the QCD-inspired 
reciprocity properties in such an exotic gauge theory. Some breaking of reciprocity is expected
since now the gauge structure is rather far from the QCD one and the physical arguments supporting 
reciprocity are missing or at least much weaker.

The analysis of~\cite{Beccaria:2009ny} shows that twist-one operators obey a four-loop
parity invariance closely related to the reciprocity discussed in this Review. 
This four-loop result for the twist-one  operators includes the leading-order wrapping
correction, computed using the Y-system formalism~\cite{Ysystem}. 
In the twist-two case, parity invariance is badly broken, although some remnants can still be seen
in the fine structure of the kernel $\cal P$.

\end{document}